\documentclass[twocolumn]{autart}    
\usepackage{enumitem}
 \usepackage{cite}
  \usepackage{graphicx}
  \usepackage{multirow}
  \usepackage{psfrag}
  \usepackage{commath}
  \usepackage{amsfonts}
  \usepackage{dirtytalk}
  \usepackage{array}
\newcolumntype{P}[1]{>{\centering\arraybackslash}p{#1}}
  \usepackage{xcolor}
  \usepackage{mathtools}
  \DeclarePairedDelimiter\ceil{\lceil}{\rceil}
 \DeclarePairedDelimiter\floor{\lfloor}{\rfloor}
  \usepackage{breqn}
  \usepackage{amsmath}
  
  \newcommand{\argmax}{\arg\max}
  \newcommand\ar{\mathrm{A}}
  \newcommand\dr{\mathrm{D}}
  \newcommand\lb{\alpha}
  \newcommand\co{c(\mathcal{G})}
  \newcommand\ca{c(\mathcal{G}^{\mathrm{A}}_{l,\alpha})}
  \newcommand\cd{c(\mathcal{G}^{\mathrm{D}}_{l,\alpha})}

  \newcommand\bab{\overline{\beta}^{\mathrm{A}}}

  \newcommand\eam{\mathcal{E}^{\mathrm{A}}_m}
  \newcommand\edm{\mathcal{E}^{\mathrm{D}}_m}
  \newcommand\go{\mathcal{G}}
  \newcommand\vo{\mathcal{V}}
  \newcommand\ga{\mathcal{G}^{\mathrm{A}}_k}
  \newcommand\gd{\mathcal{G}^{\mathrm{D}}_k}
  \newcommand\gdo{\mathcal{G}^{\mathrm{D}*}_k}

  \newcommand\eo{\mathcal{E}}
  \newcommand\ea{\mathcal{E}^{\mathrm{A}}_k}
  \newcommand\eob{\overline{\mathcal{E}}}
  \newcommand\eab{\overline{\mathcal{E}}^{\mathrm{A}}_k}
  \newcommand\eabm{\overline{\mathcal{E}}^{\mathrm{A}}_m}
  \newcommand\ed{\mathcal{E}^{\mathrm{D}}_k}

  \newcommand\ua{U^{\mathrm{A}}_l}
  \newcommand\ud{U^{\mathrm{D}}_l}

  \newcommand\uah{\hat{U}^{\mathrm{A}}_k}
  \newcommand\udh{\hat{U}^{\mathrm{D}}_k}
  \newcommand\ba{\beta^{\mathrm{A}}}
  \newcommand\bd{\beta^{\mathrm{D}}}

  \newcommand\ra{\rho^{\mathrm{A}}}
  \newcommand\rd{\rho^{\mathrm{D}}}

  \newcommand\ka{\kappa^{\mathrm{A}}}
  \newcommand\kd{\kappa^{\mathrm{D}}}
  
\newcommand\blfootnote[1]{%
  \begingroup
  \renewcommand\thefootnote{}\footnote{#1}%
  \addtocounter{footnote}{-1}%
  \endgroup
}

\begin{document}

\begin{frontmatter}
\title{A Rolling Horizon Game Considering Network Effect \\ in Cluster Forming for Dynamic Resilient \\ Multiagent Systems} 

\author[yur]{Yurid Nugraha}\ead{yurid@dsl.sc.e.titech.ac.jp},    
\author[ahmet]{Ahmet Cetinkaya}\ead{ahmet@shibaura-it.ac.jp},               
\author[yur]{Tomohisa Hayakawa}\ead{hayakawa@sc.e.titech.ac.jp},
\author[ishii]{Hideaki Ishii}\ead{ishii@c.titech.ac.jp},
\author[zhu]{Quanyan Zhu}\ead{quanyan.zhu@nyu.edu}  

\address[yur]{Department of  
Systems and Control Engineering, Tokyo Institute of Technology, Tokyo 152-8552, Japan}         
\address[ahmet]{Department of Functional Control Systems, Shibaura Institute of Technology, Tokyo, 135-8548, Japan}
\address[ishii]{Department of Computer Science, Tokyo Insitute of Technology, Yokohama 226-8502,  Japan}  
\address[zhu]{Department of Electrical and Computer Engineering, New York University, Brooklyn NY, 11201, USA}  


\begin{abstract}                         A two-player game-theoretic problem on resilient graphs in a multiagent consensus setting is formulated. An attacker is capable to disable some of the edges of the network with the objective to divide the agents into clusters by emitting jamming signals while, in response, the defender recovers some of the edges by increasing the transmission power for the communication signals. Specifically, we consider repeated games between the attacker and the defender where the optimal strategies for the two players are derived in a rolling horizon fashion based on utility functions that take both the agents' states and the sizes of clusters (known as network effect) into account. The players' actions at each discrete-time step are constrained by their energy for transmissions of the signals, with a less strict constraint for the attacker. Necessary conditions and sufficient conditions of agent consensus are derived, which are influenced by the energy constraints. The number of clusters of agents at infinite time in the face of attacks and recoveries are also characterized. Simulation results are provided to demonstrate the effects of players' actions on the cluster forming and to illustrate the players' performance for different horizon parameters.
\end{abstract}

\end{frontmatter}

\section{Introduction}
Applications of large-scale networked systems have rapidly grown in various areas of critical infrastructures including power grids and transportation systems. Such systems can be considered as multiagent systems where a number of agents capable of making local decisions interact over a network and exchange information to reach a common goal \cite{FB-LNS}. While wireless communication plays an important role for the functionality of the network, it is also prone to cyber attacks initiated by malicious adversaries \cite{sandb,booki}. 

Jamming attacks in consensus problems of multiagent systems have been studied in \cite{tesi,kikuchi,ahm2}. Noncooperative games between attackers and other players protecting the network are widely used to analyze security problems, including jamming attacks \cite{li,jia} and injection attacks \cite{pir,sanjab,ylitac}.

In a jamming attack formulation, it is natural to consider that the jammer/the attacker has an energy constraint such that, if it is not connected to energy sources, it is impossible to attack all communication links of the network at all times \cite{kikuchi,ahm}. In the context of game-theoretical approaches, this constraint becomes important to characterize the strategic behaviors of the players \cite{li}. 

When the links in the network are attacked, the agents may become disconnected from other agents, resulting in several groups of connected agents, or \textit{clusters}. The work \cite{net} proposed the notion of network effect/network externality, which refers to the utility of an agent in a certain cluster depending on how many other agents belong to that particular cluster. Such a concept has been used to analyze grouping of agents on, e.g., social networks and computer networks, as discussed in \cite{sns1,sns2}.

Rolling horizon control has been used to handle systems with uncertainties. It is also studied in the context of networked control \cite{mzhu3,mzhu4}, where there may be additional uncertainties related to communications among agents in the networks. Rolling horizon approaches are also discussed in noncooperative security game settings in \cite{mzhu,mzhu2}, where horizon lengths affect the resilience of the system. Rolling horizon approaches have also been used to handle the constraints in the system, e.g., in an agent with obstacle avoidance constraints \cite{schou,schou2}.

In this paper, we consider a security problem in a two-player game setting between an attacker, who is motivated to disrupt the communication among agents by attacking communication links, and a defender, who attempts to recover some of the attacked links. We formulate the problem based on \cite{arxivYur,chen}, which use graph connectivity to characterize the game and the players' strategies. The game in this paper is played repeatedly over discrete time in the context of multiagent consensus. 


{\color{black}As a results of these persistent attacks and recoveries, under consensus protocol cluster forming emerges among the agents of the networks with different clusters having different agents' states. Cluster forming in multiagent systems has been studied in, e.g., \cite{altafini,maria,shang}, where the relations among certain agents may be hostile. In this paper, we approach clustering from a different viewpoint based on a game-theoretic formulation. Specifically, the players of the game consider network effect/network externality\cite{net} to form clusters among agents. Their utilities are determined by how the network is disconnected into groups of agents as well as how the players’ actions affect the states of the agents at each time. Under this setting, the number and the size of the clusters are influenced by how strong the attacks are; the stronger attacker is supposed to be able to separate agents into more smaller clusters, and vice versa.

In the resilient network setting, it is common that there exists a network manager who is aware of the incoming attack, since the agents try to communicate with their neighbor agents at all time and thus quickly know if some of their neighbors do not send any signal. The network manager then tries to prepare a defense plan to quickly recover from such attacks and to repel the subsequent attacks. 

From the attacker's viewpoint, it is also common that the attacker knows which edges of the network are the most vulnerable as well as how powerful the network manager is, e.g., the manager's remaining resources. Therefore, we believe that this sequential model can be applied to several real-world settings.}

{\color{black} The main contribution of this paper is that we introduce a repeated game played repeatedly over time to model the decision making process between the attacker and the defender in the context of network security. It is then natural to explore how these games affect the networks and state evolution of the agents. Consensus protocol is considered due to its simple characterization, where all agents should converge in the case of no attack. More specifically, in comparison to \cite{arxivYur,chen}, our contribution is threefold: (i) We introduce more options for the attacker's jamming signal strengths; (ii) the game consists of multiple attack-recovery actions, resulting in more
complicated strategies; and (iii) we consider a rolling horizon approach for the players so that their strategies may be modified as they obtain new knowledge of the status of the system.}

{\color{black} More specifically, it is now possible for the attacker to disable links with stronger intensity of attack signals so that the defender is unable to recover those links (the decision on which edges are to be attacked with stronger attack signals is made at the same time as the decision on which edges are to be attacked with normal attack signals); this feature is motivated by \cite{yangj,beibei}. In practice, this is possible when the attacker emits stronger jamming signals that takes more resource that results in much lower signal-to-interference-plus-noise ratio (SINR) so that it is not possible for the defender to recover the communication on those links with its limited recovery strength.} On the other hand, we consider games consisting of multiple parts, where the players need to consider their future utilities and energy constraints when deciding their strategies at any point in time. {\color{black} This setting enables the the players to think further ahead and prioritize their long-term payoffs, compared to in a single-step case}. The players recalculate and may override their strategies as time goes on, according to the rolling horizon approach. A related formulation without rolling horizon is discussed in \cite{yur4}, where the players are not able to change their strategies decided at earlier times.

The paper is organized as follows. In Section~2, we introduce the framework for the attack-recovery sequence, cluster forming among agents, and energy consumption models of the players. The utility functions of the games in rolling horizon approach of the repeated games is discussed in Section~3, whereas the game structure is characterized in Section~4. 
In Section~5, we analyze some conditions of consensus among agents, which are related to the parameters of the underlying graph and the players' energy constraints. We continue by discussing the cluster forming of agents when consensus is not achieved in Section~6. The equilibrium characterization of the game under certain conditions is discussed in Section~7. We then provide numerical examples on consensus and cluster forming in Section~8 and conclude the paper in Section~9. The conference version of this paper appeared in \cite{yur5}, where we consider a more restricted situation on how often players update their strategies.

The notations used in this paper are fairly standard. We denote by $|\cdot|$ the cardinality of a set. The floor function and the ceiling function are denoted by $\floor{\cdot}$ and $\ceil{\cdot}$, respectively. The sets of positive and nonnegative integers are denoted by $\mathbb{N}$ and $\mathbb{N}_0$, respectively.

\vspace{-0.0cm}
\section{Attack/Recovery Characterization for Multiagent Systems Under Consensus Dynamics}\label{sec2}
\begin{figure*}[t]
\begin{minipage}[c]{0.55\textwidth}
      \hspace{-10pt}
     \centering
        \psfrag{time}{$k$}
        \psfrag{d6}{{Edge}}
        \psfrag{eza}{ }
        \psfrag{ezb}{ }
         \psfrag{del01}{\small $0$}
        \psfrag{del02}{\small $1T$}
        \psfrag{del03}{\small $2T$}
        \psfrag{del22}{ }
        \psfrag{del23}{ }
        \psfrag{del24}{ }
        \psfrag{del25}{ }
        \psfrag{ez4}{\small $l=1$}
        \psfrag{ez1}{\small $l=2$}
        \psfrag{ez3}{\small $l=3$}
        \psfrag{ez5}{\small \hspace{-0.1cm} horizon length $h$}
        \psfrag{ez6}{\small \hspace{-0.2cm} 2nd game}
        \psfrag{ez7}{ }
        \psfrag{ez8}{\small horizon length $h$}
        \psfrag{ez9}{\small game period $T$}
      \includegraphics[width=10.5 cm]{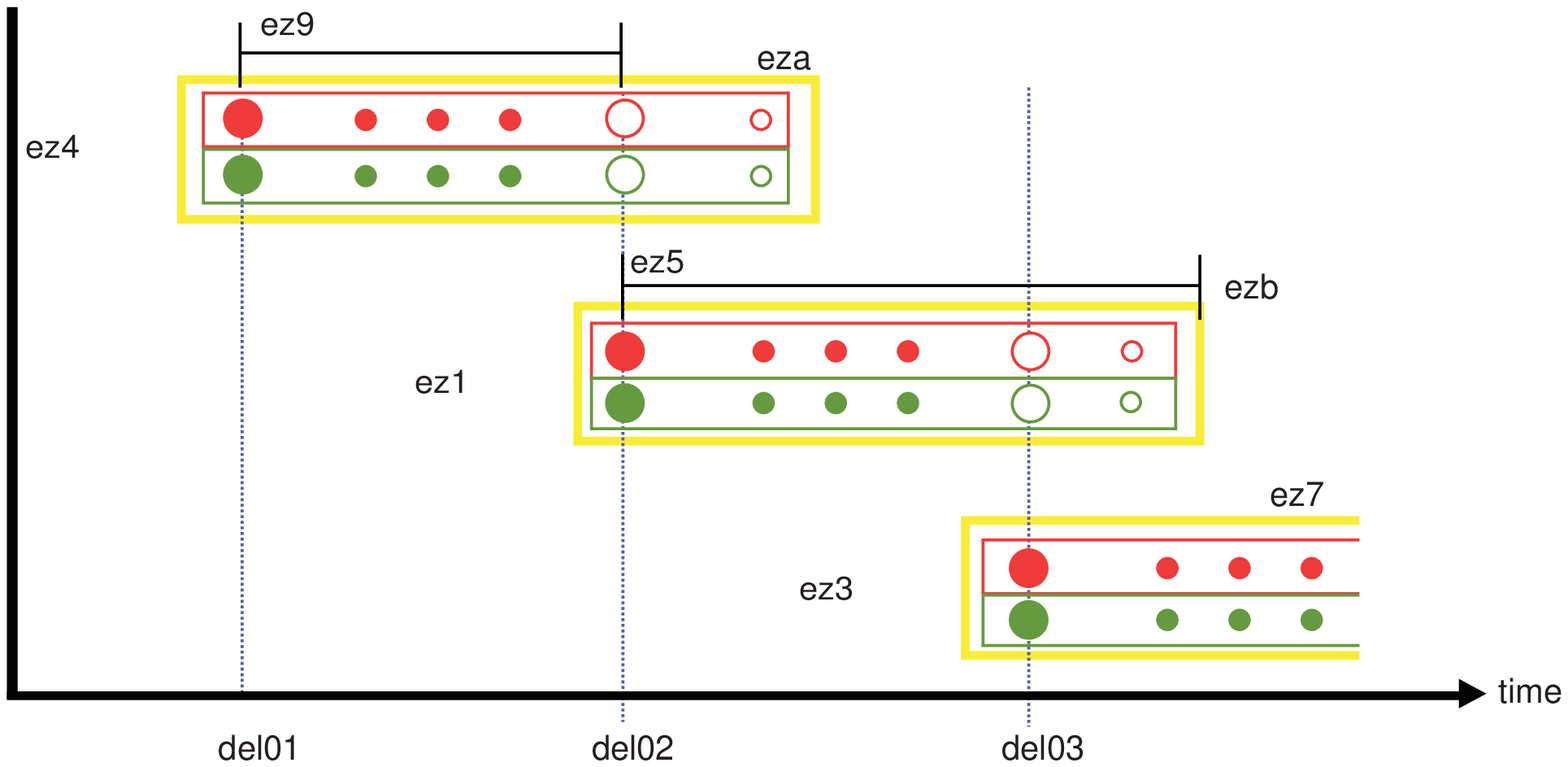}
        \vskip -5pt
        \caption{\small \textcolor{black}{Illustration of the games played over discrete time $k$ with rolling horizon approaches by the players.}}
        \label{fig:a}
        \end{minipage}
        \hfill
        \begin{minipage}[c]{0.40\textwidth}
        \centering
        \psfrag{del00}{\footnotesize $0$}
        \psfrag{del01}{\footnotesize $1$}
        \psfrag{del02}{\footnotesize $2$}
        \psfrag{del03}{\footnotesize $3$}
        \psfrag{del04}{\footnotesize $k$}
        \psfrag{del05}{\footnotesize $4$}
        \psfrag{EN}{\small Energy}
        \psfrag{EN1}{\scriptsize $\kappa^{\mathrm{A}}$}
        \psfrag{EN2}{\textcolor{white}{--} Time}
        \includegraphics[width=6 cm]{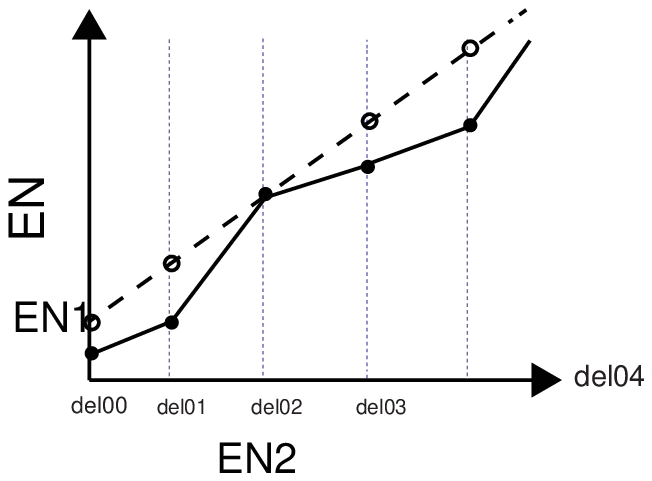}
        \vskip -5pt
        \caption{\small Energy constraint of the attacker considered in the formulation. The dashed line represents the total supplied energy to spend. The filled circles representing the actual energy consumed by the player should be below the dashed line.}
        \label{ene}
        \vskip -7pt
    \end{minipage}
    \vskip 0pt
    \end{figure*}  
 We consider a multiagent system of $n$ agents communicating to each other in discrete time in the face of jamming attacks. The agents are aiming to converge to a consensus state by interacting with each other over the communication network. The network topology for the normal operation is given by an undirected and connected graph $\go=(\vo,\eo)$. The graph consists of the set $\vo$ of vertices representing the agents and the set $\eo \subseteq \vo \times \vo$ of edges representing the communication links. The edge connectivity \cite{FB-LNS} of the connected graph $\go$ is denoted by $\lambda$.
 
 Each agent $i$ has the scalar state $x_i[k]$ following the discrete-time update rule at time $k \in \mathbb{N}_0$ given by
 \begin{align} 
        x_i[k+1]&=x_i[k]+u_i[k], \quad x[0]=x_0, \label{state} 
    \end{align}
 where $u_i[k]$ denotes the control input applied to agent~$i$. We assume that $u_i[k]$ is constructed as the weighted sum of the state differences between agent $i$ and its neighbor agents, commonly used in, e.g., \cite{dib2}, which is given by
    \begin{align} 
        u_i[k]&=\sum_{j \in {\mathcal{N}_{i}}[k]} a_{ij}(x_j[k]-x_i[k]), \label{state2}
    \end{align}
 where $\mathcal{N}_{i}[k]$ denotes the set of agents that can communicate with agent $i$ at time $k$, and $a_{ij}$ represents the weight of edge  $(i,j) \in \eo$ such that $\Sigma^n_{j=1,j \ne i} a_{ij}<1$, $i \in \vo$ to ensure that the agents achieve consensus without any attack.
 
We assume that the jamming attacks on an edge affect the communication between the two agents connected by that attacked edge. As a result, the set $\mathcal{N}_i[k]$ may change, and the resulting communication topology can be disconnected at time $k$. Such jamming attacks are represented by the removal of edges in $\go$. On the other hand, within the system there is a defender that may be capable of maintaining the communication among the agents, e.g., by asking agents to send stronger communication signals to overcome the jamming signals. This action is represented as rebuilding some of the attacked edges.

From this sequence of attacks and recoveries, we characterize the attack-recovery process as a two-player game between the \textit{attacker} and the \textit{defender} in terms of the communication links in the network. In other words, the graph characterizing the networked system is \textit{resilient} if the group of agents is able to recover from the damages caused by the attacker. However, there may be cases where the resiliency level of the graph is reduced if the jamming signals are sufficiently strong such that the defender cannot recover. Note that to achieve consensus, the agents need \emph{not} be connected for \emph{all} time.

In this paper, we consider the case where the attacker has two types of jamming signals in terms of their strength, \textit{strong} and \textit{normal}. The defender is able to recover only the edges that are attacked with normal strength. In the following subsections, we first describe the sequence of  attacks and recoveries and characterize some constraints on the players' energy and computational ability that we need to impose as well as how the objective of the problem is formulated.

\subsection{Attack-Recovery Sequence}

In our setting, at each discrete time $k$, the players (the attacker and the defender) decide to attack/recover certain edges in two stages, with the attacker acting first and then the defender. Specifically, at time $k$ the attacker attacks $\go$ by deleting the edges $\ea\subseteq \eo$ with normal jamming signals and $\eab\subseteq \eo$ with strong jamming signals with $\ea \cap \eab = \emptyset$, whereas the defender recovers $\ed \subseteq \ea$. As mentioned earlier, the defender is not able to recover the edges attacked with strong jamming signals, i.e., $\ed \cap \eab=\emptyset$. Due to the attacks and then the recoveries, the network changes from $\go$ to $\ga:=(\vo,\eo \setminus (\ea\cup \eob^{\mathrm{A}}_k))$  and further to $\gd:=(\vo,(\eo \setminus (\ea\cup \eob^{\mathrm{A}}_k)) \cup \ed)$ at time $k$. The agents then communicate to their neighbors $\mathcal{N}_i[k]$ based on this resulting graph $\gd$. 
    
    In this game, the players attempt to choose the best strategies in terms of edges attacked/recovered $(\eob^{\mathrm{A}}_k,\ea)$ and $\ed$ to maximize their own utility functions. Here, the games are played every \textcolor{black}{game period} $T$ time steps and the $l$th game is defined over the horizon of $h$ steps from time $(l-1)T$ to $(l-1)T+h-1$, with $l \in \mathbb{N}$ and $1 \leq T \leq h$. The players make decisions in a \textit{rolling horizon} fashion; the optimal strategies obtained at $(l-1)T$ for the future time may be overridden when the players recalculate their strategies at time $lT$ when the next game starts. \textcolor{black}{Fig.~\ref{fig:a} illustrates the discussed sequence over time with $h=8$ and $T=4$, where the filled circles indicate the implemented strategies and the empty circles indicate the strategies of the game that are discarded.} {\textcolor{black}{In this setting,  the \textit{horizon length} $h$ indicates the computational ability, i.e., how long in the future the players can plan their strategies, whereas the \textit{game period} $T\leq h$ indicates the players' adaptability, i.e., how long the players apply the obtained strategies without updating (shorter $T$ means that a player is more adaptable).}} The rolling horizon game structure will be discussed in Section~\ref{satD} in more detail.
    
    \subsection{Energy Constraints}
    The actions of the attacker and the defender are affected by the constraints on their energy resources. It is assumed that the total supplied energy for the players increases linearly in time; furthermore, the energy consumed by the players is proportional to the number of attacked/recovered edges. Here we suppose that the players initially possess certain amount of energy $\ka$ and $\kd$ for the attacker and the defender, respectively. Moreover, the players are assumed to be able to supply energy wirelessly to devices that obstruct/retain communication signals between the agents so that the energy supply rates to these devices are limited by the constant values of $\ra$ and $\rd$ every discrete time step. {\color{black} These devices are supposed to have unlimited battery capacity and thus can be supplied constantly by the players with a linear rate $\ra$ or $\rd$.}
    
    For the attacker, the strong attacks on $\eab$ take $\bab>0$ energy per edge per unit time whereas the normal attacks on $\ea$ take $\ba>0$ cost per edge, with $\bab>\ba$. The total energy used by the attacker is constrained as
    \begin{align}
    \sum_{m=0}^{k} & (\bab|\eabm|+\ba|\eam|) \leq \ka + \ra k \label{a}
    \end{align} 
     for any time $k$, where $\kappa^{\mathrm{A}}\geq \ra>0$. This implies that the total energy spent by the attacker cannot exceed the available energy characterized as the sum of the initial energy $\kappa^{\mathrm{A}}$ and the supplied energy $\rho^{\mathrm{A}}k$ by time $k$. This energy constraint restricts the number of edges that the attacker can attack. Note that the attacker's available energy increases by $\ra$ at each $k$. The condition $\ka \geq \ra$ allows the attacker to have at least the same attack ability at time $k=0$.

    Fig.~\ref{ene} illustrates the energy constraint of the attacker, where the dashed line with slope $\ra$ represents the total supplied energy and the filled circles indicate the total energy spent. A critical case is when $\ba<\ra$, since it is possible for the attacker to attack at least one edge for all times. This will have implications on the consensus and cluster forming of the agents, as we will discuss later.
    
    The energy constraint for the defender is similar to (\ref{a}):
    \begin{equation} \label{en.d}
        \sum_{m=0}^{k} \bd|\edm| \leq \kd + \rd k,
    \end{equation}
    with $\kd\geq \rd>0$ and $\bd>0$. Note that there is a single term on the left-hand side because there is only one type of recovery signals for the agents.
    
    \section{Utility Functions with Cluster Forming and Agent-group Index Considerations}\label{sec3}
    
    In our game setting, the attacker tries to make the graph disconnected to separate the agents into clusters. Here, we introduce a few notions related to grouping/clustering of agents. In a given subgraph $\go'=(\vo,\eo')$ of $\go$, the agents may be divided into $\overline{n}(\go')$ number of \textit{groups}, with the groups $\vo'_1,\vo'_2,\ldots,\vo'_{\overline{n}(\go')}$ being a partition of $\vo$ with $\cup_{p=1}^{\overline{n}(\go')} \vo'_p = \vo$ and $\vo'_p \cap \vo'_q = \emptyset$, if $p \ne q$. There is no edge connecting different groups, i.e., $e_{i^\prime,j^{\prime}}\notin \eo^{\prime}, \forall i^{\prime}\in \vo^{\prime}_p, j^{\prime} \in \vo^{\prime}_q$. We also call each subset of agents taking the same state at infinite time as a \textit{cluster}, i.e., $\lim_{k \to \infty} (x_{i}[k]- x_{j}[k])=0$ implies that agents $i$ and $j$ belong to the same cluster.
    
    In the considered game, the attacker and the defender are concerned about the number of agents in each group. Specifically, we follow the notion of \textit{network effect/network externality}\cite{net}, where the utility of an agent in a certain group depends on how many other agents belong to that particular group. In the context of this game, the attacker wants to isolate agents so that fewer agents are in each group, while the defender wants as many agents as possible in the same group. We then represent the level of grouping in the graph $\go'$ by the function $c(\cdot)$, which we call the \textit{agent-group index}, given by 
    \begin{align}\label{cluster}
    c(\go'):=\sum_{p=1}^{\overline{n}(\go')} |\vo'_p|^2 -|\vo|^2 \quad (\leq 0).
    \end{align}
    
\begin{figure}[t]
        \centering
        \psfrag{e1}{\text{(a)} $\go_{\mathrm{A}}$}
        \psfrag{e2}{\text{(b)} $\go_{\mathrm{B}}$}
        \psfrag{e3}{\text{(c)} $\go_{\mathrm{C}}$}
        \psfrag{e4}{\text{(d)} $\go_{\dr}$}
        \includegraphics[width=8 cm]{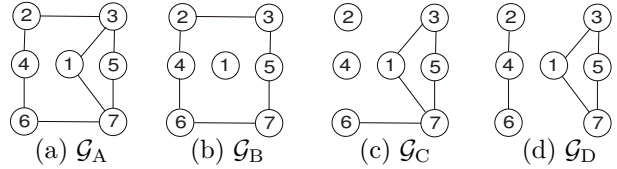}
        \vskip -5pt
        \caption{Graphs and their agent-group indices: (a)  $c(\go_{\mathrm{A}})=0$, (b) $c(\go_{\mathrm{B}})=-12$, (c)  $c(\go_{\mathrm{C}})=-22$, and (d) $c(\go_{\dr})=-24$. Note that $c(\go_{\mathrm{C}})$ is larger than $c(\go_{\dr})$, even with more number of groups.}
        \label{fig:clus}
    \end{figure}

    The value of $c(\go')$ is 0 if $\go'$ is connected, since there is only one group (i.e., $\overline{n}(\go')=1$). A larger value (closer to 0) of $c(\go')$ implies that there are fewer groups in graph $\go'$, and/or each group has more agents. The agent-group indices of some graphs are shown in Fig. \ref{fig:clus}. Here, it is interesting that $c(\go_{\dr})$ is smaller than $c(\go_{\mathrm{C}})$, even though $\go_{\mathrm{C}}$ has more groups. It is because the largest cluster is constituted by more agents in $\go_{\mathrm{C}}$ than the case of $\go_{\dr}$. Thus, for an attacker who tries to reduce the number of agents in one cluster, $\go_{\dr}$ is preferable to $\go_{\mathrm{C}}$.
    
    {\color{black} In our problem setting, the players also consider the effects of their actions on the agent states when attacking/recovering.} For example, the attacker may want to separate agents having state values with more differences in different groups. We specify the agents' state difference $z_k$ as
    \begin{align}
    z_k(\eab,\ea,\ed):= x^{\mathrm{T}}[k+1] L_{\mathrm{c}} x[k+1], \label{z}
    \end{align}
    with $L_{\mathrm{c}}$, for simplicity, being the Laplacian matrix of the complete graph with $n$ agents. That is, (\ref{z}) represents the sum of squares of the state differences of all the agent pairs. {\color{black} This implies that all state differences between any pair of agents are worth the same and thus the players do not prioritize any connection between agents.} 
    
    The attacked and recovered edges $(\eab,\ea,\ed)$ will affect $x[k+1]$ in accordance with (\ref{state}) and (\ref{state2}), and in turn the value of $z_k$. Note that the value of $z_k$ is nonincreasing over time \cite{FB-LNS} even if some agents are left disconnected from other agents under attacks. {\color{black} This sum-of-square characterization of the agents' state difference is commonly used and essentially the same to our previous work \cite{yur4} for the continuous-time setting; here, we extend the formulation to comply with the discrete-time setting by considering the states at one time step ahead $k+1$.} 
    
    Now, we combine the two measures in (\ref{cluster}) and (\ref{z}) to construct the utility functions for the game in a zero-sum manner. Specifically, for the $l$th game starting at time $k=(l-1)T$, the attacker and the defender's utility functions take account of the agent-group index $c(\cdot)$ and the difference $z_k$ of agents' states over $h$ horizon length from time $(l-1)T$ to $(l-1)T+h-1$. With weights $a,b \geq 0$, the utilities for the $l$th game $\ua$ for the attacker and $\ud$ for the defender are, respectively, defined by
    \begin{align}
    \ua  & := \textcolor{black}{\sum_{k=(l-1)T}^{(l-1)T+h-1} (a z_k - bc(\go^{\dr}_k)) ,} \label{ua} \\
    \ud  & := -\ua. \label{ud}
    \end{align}
    In our setting both players attempt to maximize their utilities at the start of each game $l$. \textcolor{black}{The values of $a$ and $b$ represent the preference of the players towards either a long-term agent clustering or a short-term agent-grouping. A higher value of $a$ implies that the players prefer to focus on the agent states and the subsequent cluster forming, whereas a higher value of $b$ implies that they focus on the agent-grouping more. We suppose that both players know the underlying topology $\go$ as well as the states of all agents $x_i[k]$.}
    
    \section{Rolling Horizon Game Structure} \label{satD}
    We are interested in finding the subgame perfect equilibrium \cite{fudenberg} of this game outlined in Section~\ref{sec3}. To this end, the game is divided into some subgames/decision-making points. The subgame perfect equilibrium must be an equilibrium in every subgame. The optimal strategy of each player is obtained by using a backward induction approach, i.e., by finding the equilibrium from the smallest subgames. The tie-break condition happens when the players' strategies result in the same utility. In this case, we suppose that the players choose to save their energy by attacking/recovering less edges unless they have enough energy to attack/recover all edges in every subsequent steps, in which case they attack/recover more edges.

    Due to the nature of the rolling horizon approach, the strategies obtained from the $l$th game, i.e., attacked and recovered edges, are applied only from time $(l-1)T$ to $lT-1$. Specifically, in the $l$th game for time $(l-1)T$ to $(l-1)T+h-1$, the strategies of both players are denoted by $((\eob^{\mathrm{A}}_{l,1},{\eo}^{\mathrm{A}}_{l,1},{\eo}^{\dr}_{l,1}),\ldots,(\eob^{\mathrm{A}}_{l,h},{\eo}^{\mathrm{A}}_{l,h},{\eo}^{\dr}_{l,h}) )$, with $(\eob^{\mathrm{A}}_{l,\lb},\eo^{\mathrm{A}}_{l,\lb},\eo^{\dr}_{l,{\lb}})$ indicating the strategies at the $\lb$th step of the $l$th game with $\lb \in \{1,\ldots,h\}$. Note that here we show the strategies with two subscripts representing the game and the step indices along the time axis. From the above set of strategies, only $((\eob^{\mathrm{A}}_{l,1},{\eo}^{\mathrm{A}}_{l,1},{\eo}^{\dr}_{l,1}),\ldots,(\eob^{\mathrm{A}}_{l,T},{\eo}^{\mathrm{A}}_{l,T},{\eo}^{\dr}_{l,T}))$ is applied. Recall that $h$ is taken to be greater than or equal to $T$. Therefore, for the $l$th game from time $(l-1)T$ to $lT-1$, the strategy applied will be written as $((\eob^{\mathrm{A}}_{(l-1)T},{\eo}^{\mathrm{A}}_{(l-1)T},{\eo}^{\dr}_{(l-1)T}),\ldots,(\eob^{\mathrm{A}}_{lT-1},{\eo}^{\mathrm{A}}_{lT-1},{\eo}^{\dr}_{lT-1})):=((\eob^{\mathrm{A}}_{l,1},{\eo}^{\mathrm{A}}_{l,1},{\eo}^{\dr}_{l,1}),\ldots,(\eob^{\mathrm{A}}_{l,T},{\eo}^{\mathrm{A}}_{l,T},{\eo}^{\dr}_{l,T}))$.
    
    We look at how the optimal edges can be found by an example with $h=2$ and $T=1$ or $2$. In this case, for the $l$th game over time $(l-1)T$ and $(l-1)T+1$, the optimal strategies of the players are given by
    \begin{align}
    \eo^{\dr*}_{l,2} (\eob^{\mathrm{A}}_{l,2},\eo^{\mathrm{A}}_{l,2}) & \in \argmax_{\eo^{\dr}_{l,2}}  U^{\dr}_{l,2}, \label{st1} \\
    (\eob^{\mathrm{A}*}_{l,2}(\eo^{\dr}_{l,1}),\eo^{\mathrm{A}*}_{l,2}(\eo^{\dr}_{l,1})) & \in \argmax_{(\eob^{\mathrm{A}}_{l,2},\eo^{\mathrm{A}}_{l,2})} U^{\mathrm{A}}_{l,2}, \label{st2} \\
     \eo^{\dr*}_{l,1}(\eob^{\mathrm{A}}_{l,1},\eo^{\mathrm{A}}_{l,1}) & \in \argmax_{\eo^{\dr}_{l,1}} U^{\dr}_l, \label{s1} \\
     (\eob^{\mathrm{A}*}_{l,1},\eo^{\mathrm{A}*}_{l,1}) & \in \argmax_{(\eob^{\mathrm{A}}_{l,1},\eo^{\mathrm{A}}_{l,1})} U^{\mathrm{A}}_l, \label{s2}
    \end{align}
    where $U^{\mathrm{A}}_{l,\lb}$ and $U^{\dr}_{l,\lb}$ are defined as parts of $\ua$ and $\ud$, respectively, calculated from the $\lb$th step to the last ($h$th) step of the $l$th game, i.e., $U_{l,\lb}^{\ar} = - U_{l,\lb}^{\dr} := \sum_{(l-1)T+\alpha-1}^{(l-1)T+h-1} (az_k-bc(\gd))$. In this case with $h=2$, the functions $U^{\mathrm{A}}_{l,2}$ and $U^{\dr}_{l,2}$ are based on the values of $az_k$ and $b\go^\dr_k$ at $k=(l-1)T+1$ only. Note that to find $(\eob^{\mathrm{A}*}_{l,1},\eo^{\mathrm{A}*}_{l,1})$, one needs to obtain $\eo^{\dr*}_{l,1}(\eob^{\mathrm{A}}_{l,1},\eo^{\mathrm{A}}_{l,1})$ beforehand. Likewise, to find $\eo^{\dr*}_{l,1}$, one needs to obtain $(\eob^{\mathrm{A}*}_{l,2}(\eo^{\dr}_{l,1}),\eo^{\mathrm{A}*}_{l,2}(\eo^{\dr}_{l,1}))$. Similarly, to find $(\eob^{\mathrm{A}*}_{l,2},\eo^{\mathrm{A}*}_{l,2})$, the edges $\eo^{\dr*}_{l,2}(\eob^{\mathrm{A}}_{l,2},\eo^{\mathrm{A}}_{l,2})$ must be obtained beforehand. {\textcolor{black}{Note that deriving the optimal strategies above is subject to the energy constraints (\ref{a}) and (\ref{en.d}).}}
    
    For $h>2$, the players' optimal strategies consist of $2h$ parts similar to those in (\ref{st1})--(\ref{s2}), with one time step consisting of two parts of strategies corresponding to the number of players. They are solved by the players at every time $k=(l-1)T$ of the $l$th game, $l \in \mathbb{N}$. With $T=h$, the players do not have chance to override their strategies, which removes the rolling horizon aspect of the game.
    
    We will find the optimal strategies of the players by computing all possible combinations, since the choices of edges are finite. {\textcolor{black}{From the optimization problems specified above, the players examine at most $3^{|\eo|}2^{|\eo|}h$ number of combinations of attacked and recovered edges for utility evaluations, since they have to foresee the opponent's response as well. Note that the attacker has three possible actions on an edge: no attack, attack with normal signals, and attack with strong signals, whereas the defender has only two actions: recover or not recover. Here we can see that the number of computation increases exponentially with respect to the number of edges in the underlying graph. To address scalability issues, we may find edges that are easier to attack first, i.e., edges that result in the formation of new groups if attacked, and limit the strategy choices over those edges only.}}
    
    Our previous works \cite{arxivYur,yur4} considered related games in continuous time, where the timings for launching attack/defense actions are also part of the decision variables. This aspect complicated the formulation, making it difficult to study games over a time horizon. In this paper, we simplify the timing issue and instead introduce the rolling horizon feature. This enables the players to consider the cluster forming in a longer time range, which is especially important when consensus among agents is obstructed by adversaries.
    
    {\color{black} With this rolling horizon setting, it is important for a player to know what the opponent's previous action at the previous step of the game is in order to know its position at the game tree, i.e., which subgame is the player's playing. For example, if the defender does not know which edges are previously attacked, then it cannot properly calculate the value of the utility function (\ref{ud}).}
    
    \section{Consensus Analysis}\label{thre}
    In this section, we examine the effect of the game structure and players' energy constraints on consensus. 
    
    We will begin the analysis by looking at the case of certain energy conditions of the players. Specifically, if a player has enough energy to attack/recover all edges from a certain step of the game, then it will use all of their energy to attack/recover as many edges as they can in the subsequent steps. We will confirm this point formally in the following. For simplicity, we denote the total energy that the defender consumed before the $l$th game as $\tilde{\beta}^{\dr}_l:=\sum_{k=0}^{(l-1)T-1} \bd|\ed|$ and the total energy that the defender may consume from the $1$st to the $\lb$th step of the $l$th game as $\hat{\beta}^{\dr}_{\lb}:=\sum_{m=1}^{\lb} \bd|\eo^{\dr}_{l,m}|$, where we omit the index $l$ from the left-hand side, with a slight abuse of notation. Similarly, for the attacker we denote $\tilde{\beta}^{\ar}_l:=\sum_{m=0}^{(l-1)T-1}( \ba|\eam|+\bab|\eabm|)$ and  $\hat{\beta}^{\ar}_{\lb}:=\sum_{m=1}^{\lb} (\ba|\eo^{\mathrm{A}}_{l,m}|+\bab|\eob^{\mathrm{A}}_{l,m}|)$.
    
    We discuss in Lemma~\ref{lem1}~(resp., Lemma~\ref{lem2}) the optimal strategy of the defender (resp., attacker) at the $\alpha$th step of the game given certain energy conditions mentioned in Section~\ref{sec2}. This characterization of optimal strategy of the defender (resp., attacker) will be useful to obtain the necessary (resp., sufficient) conditions for consensus not to happen. 
    
    \subsection{Necessary Conditions for not Reaching Consensus}
    This subsection discusses necessary conditions for the agents to be separated into different clusters for infinitely long duration without achieving overall consensus. We first discuss the defender's optimal strategy on some games with specific conditions in Lemmas~\ref{lem1}~and~\ref{lem2}. In Lemma~\ref{lem1}, we state the defender's optimal strategy at any step of the $l$th game given a  certain energy condition.
    
    \begin{lem} \label{lem1}
    If the defender's total energy $\tilde{\beta}^{\dr}_{l}+\hat{\beta}^{\dr}_{\hat{\lb}-1}$ consumed before the $\hat{\lb}$th step of the $l$th game satisfies 
     \begin{align}
     \tilde{\beta}^{\dr}_{l}+\hat{\beta}^{\dr}_{\hat{\lb}-1} &  \notag \\ \leq \kd + \rd & ((l-1)T+\hat{\lb}-1)  - (h-\hat{\lb}+1)|\eo|\bd, \label{en.d1}
    \end{align}
     then $\eo^{\dr*}_{l,\lb}=\eo^{\mathrm{A}*}_{l,\lb}$ for all $\lb\geq\hat{\lb}$, i.e., the defender will recover all normally attacked edges from the $\hat{\lb}$th step.
    \end{lem}
    \begin{pf}
    We first look at the last ($h$th) step of the $l$th game. Since the game consists of a horizon of $h$ steps, the last step of the game corresponds to the last decision-making point, in which the players' strategies cannot influence the decision already made in the previous steps of the same game. Hence, in the last step of the $l$th game the players do not save their energy by attacking/recovering less edges. 
    
    From the defender's energy constraint (\ref{en.d}), it is clear that at any time $k$, the set of edges that the defender recovers is bounded as $|\eo^{\dr}_k|\leq {\frac{\kd + \rd k-\sum_{m=0}^{k-1} \bd|\edm|}{\bd}}$. Thus, at the $h$th step, recovered edges satisfy $|\eo^{\dr}_{l,h}|\leq |\eo^{\dr\prime}_{l,h}|$ with $|\eo^{\dr\prime}_{l,h}|:= \min\{\floor{\frac{\kd + \rd ((l-1)T+h-1)-(\tilde{\beta}^{\dr}_{l}+\hat{\beta}^{\dr}_{h-1})}{\bd}},$ $|\eo^{\ar*}_{l,h}|\}$. 
    
    
    Depending on which edges are normally attacked, the defender may not recover the maximum number $|\eo^{\dr\prime}_{l,h}|$ of edges. If the defender's optimal strategy given normally attacked edges $\eo^{\ar}_{l,h}$ is not to recover $|\eo^{\dr\prime}_{l,h}|$ number of edges, i.e., recover less, then the defender will be able to obtain more utility $U^{\dr}_{l,h}(\eob^{\ar}_{l,h},\eo^{\ar}_{l,h},\eo^{\dr}_{l,h})>U^{\dr}_{l,h}(\eob^{\ar}_{l,h},\eo^{\ar}_{l,h},\eo^{\dr\prime}_{l,h})$. However, under (\ref{en.d1}) with $\alpha=h$ the defender has sufficiently high energy, and thus the utility becomes $U^{\dr}_{l,h}(\eob^{\ar}_{l,h},\eo^{\ar}_{l,h},\eo^{\dr}_{l,h})>U^{\dr}_{l,h}(\eob^{\ar}_{l,h},\eo^{\ar}_{l,h},\eo^{\ar}_{l,h})=U^{\dr}_{l,h}(\eob^{\ar}_{l,h},\emptyset,\emptyset)$. 
    It then follows that as long as the defender has enough energy, it will recover all optimal edges attacked normally at the $h$th step, i.e., $\eo^{\dr*}_{l,h}=\eo^{\mathrm{A}*}_{l,h}$. 
    
   Next, we investigate the effect of this property on the earlier steps of the $l$th game. Since the defender's strategy at the $h$th step is not affected by its strategy at the previous (i.e., ($h-1$)th) step when $\kd + \rd ((l-1)T+h$ $-1)-(\tilde{\beta}^{\dr}_{l}+\hat{\beta}^{\dr}_{h-1}) \geq \bd|\eo|$, here the defender does not need to recover fewer edges at the $(h-1)$th step to save energy; this is because it already has enough energy to recover $\eo^{\ar*}_{l,h}$ at the $h$th step.

    Now, we derive that if $\kd + \rd ((l-1)T+h-2)-$  $(\tilde{\beta}^{\dr}_{l}+\hat{\beta}^{\dr}_{h-2})\geq 2\bd|\eo|$ at the $(h-1)$th step, then the defender will also recover $\eo^{\dr*}_{l,h-1}=\eo^{\mathrm{A}*}_{l,h-1}$. To recover all attacked edges at steps $\lb\geq\hat{\lb}$, it is then sufficient that the defender's energy
    satisfies (\ref{en.d1}) so that $\kd + \rd ((l-1)T+\lb-1) \geq \tilde{\beta}^{\dr}_{l}+\hat{\beta}^{\dr}_{\lb-1}+\bd|\eo|$, i.e., the worst-case scenario of the energy constraint (\ref{en.d}) when the defender recovers all edges, is always satisfied when $\lb\geq\hat{\lb}$.     $\hfill \square$
    
    \end{pf}
    
    From the proof above, note that if the defender's strategy is \textit{not} to recover all normally attacked edges given even if (\ref{en.d1}) is satisfied, i.e., $\eo^{\ar}_{l,\alpha}=\hat{\eo}^{\ar}\ne\eo^{\dr}_{l,\alpha}$, then the attacker will not attack $\hat{\eo}^{\ar}$ set of edges in the first place. This is because by attacking $\hat{\eo}^{\ar}$ (and considering $\eo^{\dr}_{l,\alpha}\ne \hat{\eo}^{\ar}$) the attacker's utility for step $\lb\geq\hat{\lb}$ becomes $U^{\ar}_{l,\alpha}(\cdot,\hat{\eo}^{\ar},\eo^{\dr}_{l,\alpha}\ne\hat{\eo}^{\ar})<U^{\ar}_{l,\alpha}(\cdot,\emptyset,\emptyset)$, since $U^{\dr}_{l,\alpha}(\cdot,\hat{\eo}^{\ar},\eo^{\dr}_{l,\alpha}\ne\hat{\eo}^{\ar})>U^{\dr}_{l,\alpha}(\cdot,\emptyset,\emptyset)=U^{\dr}_{l,\alpha}(\cdot,\hat{\eo}^{\ar},\hat{\eo}^{\ar})$ and $\ud=-\ua$.
    
    We also remark that in order to derive the same optimal strategy for the defender the quantity $(h-\lb+1)|\eo|$ in the right-hand side of inequality (\ref{en.d1}) can be relaxed to the maximum number of edges that the attacker can attack from step $\hat{\lb}$ to step $h$ given its energy condition. However, this number of edges may change every game, making the inequality complicated to express.
    
    Lemma~\ref{lem2} gives an interval over which, at least once, either not attacking with normal signals or recovering nonzero edges is optimal.
    \begin{lem} \label{lem2}
     There is at least one occurrence of either $\eo^{\dr}_k \ne \emptyset$ or $\eo^{\mathrm{A}}_k = \emptyset$ every $\ceil{\frac{h|\eo|\bd-\rd}{\rd T}+1}$ time steps.
    \end{lem}
    \begin{pf}
    It follows from Lemma~\ref{lem1} that in a game with index $l^{\prime}$ where (\ref{en.d1}) is satisfied for $\lb=1$, the defender always recovers edges that are attacked normally in the 1st step, i.e., $\eo^{\dr}_{l^{\prime},1}\ne\emptyset$ if $\eo^{\mathrm{A}}_{l^{\prime},1}\ne\emptyset$. We then investigate in which game inequality (\ref{en.d1}) is satisfied for $\lb=1$. Since the defender gains $\rd$ every time $k$, if $\eo^{\dr}_k=\emptyset$ for any $k\in\{0,\ldots,(l^{\prime}-1)T-1\}$, then (\ref{en.d1}) at the first step of the $l^{\prime}$th game can be written as
   $\frac{\kd + \rd (l^{\prime}-1)T}{\bd}\leq h |\eo|.$
    With $\kd=\rd$ as a worst-case scenario, the left-hand side becomes $\frac{\rd (1+(l^{\prime}-1)T)}{\bd}$, and we then obtain $l^{\prime}\geq \ceil{\frac{h|\eo|\bd-\rd}{\rd T}+1}$. 
    
    Note that the above fact holds when the defender does not recover any edge for any $k\in\{(j-1)(l^{\prime}-1)T,\ldots,j(l^{\prime}-1)T-1\}, j\in \mathbb{N}$. If the defender recovers one or more attacked edges at any $k\in\{0,\ldots,(l^{\prime}-1)T-1\}$, then the above result may not hold, i.e., the defender may not be able to recover all $\eo^{\mathrm{A}}_{l^{\prime}}$. However, it follows that during time $k\in\{(j-1)(l^{\prime}-1)T,\ldots,j(l^{\prime}-1)T-1\}$, either 1) the defender recovers nonzero edges ($\eo^{\dr}_k \ne \emptyset$), or 2) the attacker attacks no edges with normal signals ($\eo^{\mathrm{A}}_k = \emptyset$) at least once. $\hfill \square$
    \end{pf}
  
    Lemmas~\ref{lem1} and \ref{lem2} above imply that the defender is guaranteed to make recoveries from normal attacks every certain interval. Hence, the attacker needs to attack some edges strongly to prevent the recovery in order to separate agents into different clusters, as we discuss next.
    
    The following two results provide necessary conditions for consensus not to take place. We consider a more general condition in Proposition~\ref{lem3.4}, whereas in Theorem~\ref{lem3} we consider a more specific situation for the utility functions that leads to a tighter condition. Recall that $\lambda$ represents the connectivity of $\go$.
    
    {\color{black}\begin{prop} \label{lem3.4}
      A necessary condition for consensus not to happen is \textcolor{black}{$\floor{\ra/\ba} \geq \lambda$}.
    \end{prop}
    \begin{pf}
    {\color{black} In deriving this necessary condition, we suppose that there is no recovery by the defender at any time $k$. Without any recovery from the defender $(\eo^{\dr}_k=\emptyset)$, the attacker must attack at least $\lambda$ number of edges with normal signals (which take less energy) at any time $k$ to make $\go^{\dr}_k$ disconnected at all times. Otherwise, there will be time steps where the graph $\go^{\dr}_k$ is connected, which implies that consensus will still be reached.
    
    If the attacker attacks $\lambda$ edges with normal jamming signals at all times, the energy constraint (\ref{a}) becomes $(\ba \lambda-\ra) k \leq \ka$. Thus, the condition $\ra/\ba\geq \lambda$ has to be satisfied to ensure that the attacker can make $\go^{\dr}_k$ disconnected for all $k$. Note that, if the attacker does not have enough energy to disconnect $\go^{\dr}_k$ given no recovery, then it definitely cannot disconnect $\go^{\dr}_k$ in the face of recovery by the defender. $\hfill \square$}
    \end{pf}}
  \vspace{0.0cm}
  
  We now limit the class of utility functions in (\ref{ua}), (\ref{ud}) to the case of $b=0$ in the weights. This means that the players do not take account of the agent-group index in the graph, but only the states in consensus. In this case, the attacker may need more energy to prevent consensus as shown in the next theorem.
  \begin{thm} \label{lem3}
      Suppose that $b=0$. A necessary condition for consensus not to happen is $\ra/\bab \geq \lambda$.
    \end{thm}
    \begin{pf}
    We prove by contrapositive; especially, we prove that consensus always happens if $\ra/\bab<\lambda$.
    
    We first suppose that the attacker attempts to attack $\lambda$ edges strongly at all times to disconnect the graph $\gd$. From (\ref{a}), the energy constraint of the attacker at time $k$ becomes
    $(\bab \lambda-\ra) k \leq \ka$.
    This inequality is not satisfied for sufficiently large $k$ if $\ra/\bab < \lambda$, since $\bab \lambda-\ra$ becomes positive and $\ka$ is finite. Therefore, the attacker cannot attack $\lambda$ edges strongly at all times if $\ra/\bab < \lambda$, and is forced to disconnect the graph by attacking with normal jamming signals instead.
    
    Next, by Lemma~\ref{lem2} above, we show that there exists an interval of time where the defender always recovers if there are edges attacked normally, i.e., $\eo^{\dr}_{l^{\prime}}\neq \emptyset$ is optimal given that $\eo^{\mathrm{A}}_{l^{\prime}} \ne \emptyset$.

     From the definitions in (\ref{ua}), (\ref{ud}), given that $b=0$, we can see that the defender obtains a higher utility if the agents are closer. This means that given a nonzero number of edges to recover (at time $jl^{\prime}T$ described above), the defender recovers the edges connecting further agents. Specifically, for some $i\in \mathbb{N}$, for interval $[jl^{\prime}T,(j+i)l^{\prime}T]$, there is a time step where  $\ud(\eo^{\dr}_k=\eo_1) \geq \ud(\eo_2)$, with edges $\eo_1$ connecting agents with further states than agents connected by $\eo_2$. This fact implies that when recovering, the defender always chooses the further disconnected agents. Since by communicating with the consensus protocol as in (\ref{state}) the agents' states are getting closer, the defender will choose different edges to recover if the states of agents connected by recovered edges $\ed$ become close enough. Consequently, if $\ra/\bab < \lambda$, then there exists $i\in \mathbb{N}$ where the union of graphs, i.e., the graph having the union of the edges of each graph $(\vo,\bigcup((\eo\setminus(\eab \cup \ea))\cup\ed))$ over the time interval $[j(l^{\prime}-1)T,(j+i)(l^{\prime}-1)T]$, becomes a connected graph, where $l^{\prime}=\ceil{\frac{h|\eo|\bd-\rd}{\rd T}+1}$ as in Lemma~\ref{lem2} above. These intervals $[j(l^{\prime}-1)T,(j+i)(l^{\prime}-1)T]$ occur infinitely many times, since the defender's energy bound keeps increasing over time.

    It is shown in \cite{ren} that with protocol (\ref{state}), the agents achieve consensus in the time-varying graph as long as the union of the graphs over bounded time intervals is a connected graph. This implies that consensus is achieved if $(\vo,\bigcup((\eo\setminus(\eab \cup \ea))\cup\ed))$ is connected over $[l^{\prime}_i,l^{\prime}_i+1,\ldots,l^{\prime}_{i+j}]$. Thus, if $\ra/\bab < \lambda$ then consensus is achieved. $\hfill \square$
    \end{pf}
    \vspace{0.0cm}
    
    The result in Theorem~\ref{lem3} only holds for $b=0$, since with $b>0$ the defender may choose to recover the edges connecting agents that already have similar states to maximize $c(\gd)$ (instead of those connecting further agents). In such a case, the network may remain disconnected and thus the agents may converge to different states. As we see from these results, the weight values affect the necessary conditions to prevent consensus, whereas the effect of the weights on the sufficient condition (discussed later) is less straightforward. The effect of the values of $a$ and $b$ on consensus is illustrated in Section~\ref{sec6}. 
    
    \subsection{Sufficient Condition to Prevent Consensus}
    The next result provides a sufficient condition for preventing consensus. It shows that the attacker can prevent consensus if it has sufficiently large recharge rate $\ra$ given the network topology $\go$. We first state Lemma~\ref{lem1b} about the attacker's optimal strategy under some energy conditions, similar to the discussion on the defender's case above. 
    
    \begin{lem} \label{lem1b}
    The attacker's optimal strategy is $\eob^{\ar*}_{l,\alpha}=\eo$ if
    \begin{itemize}
        \item the attacker's recharge rate satisfies $\ra/\bab\geq |\eo|$, or
        \item the attacker's total energy $\tilde{\beta}^{\ar}_{l}+\hat{\beta}^{\ar}_{\lb-1}$ that it consumes before $\lb$th step of the $l$th game satisfies 
    \begin{align}
     &\tilde{\beta}^{\ar}_{l}+\hat{\beta}^{\ar}_{\lb-1} \notag \\
     & \leq  \ka + \ra ((l-1)T+\lb-1) -(h-\lb+1)\bab|\eo|. \label{lem1b1}
    \end{align}
    \end{itemize}
    \end{lem}
    \begin{pf}
    We first observe that in the $h$th step of the $l$th game the attacker does not save their energy by attacking fewer edges. 
   Since $z_{l,h}(\eo,\emptyset,$ $\emptyset)>z_{l,h}(\eob^{\mathrm{A}}_{l,h},\eo^{\ar}_{l,h},\eo^{\dr}_{l,h})$ and $c((\vo,\emptyset))\geq c((\vo,(\eo \setminus (\eob^{\ar}_{l,h}\cup \eo^{\ar}_{l,h}) \cup \eo^{\dr}_{l,h})))$ are always satisfied for any edges $\eob^{\ar}_{l,h},\eo^{\ar}_{l,h},\eo^{\dr}_{l,h}$, the function $U^{\mathrm{A}}_{h}$ always has the highest value if the attacker strongly attacks all edges $\eo$.
   It then follows that the attacker with enough energy, i.e., $\ka + \ra ((l-1)T+h-1)-(\tilde{\beta}^{\ar}_{l}+\hat{\beta}^{\ar}_{h-1})\geq \bab|\eo|$ is satisfied, will choose to attack all edges with strong signals.
    
   Similar to the proof in Lemma~\ref{lem1}, inequalities $z_{l,\alpha}(\eo,\emptyset,$ $\emptyset)>z_{l,\alpha}(\eob^{\mathrm{A}}_{l,\alpha},\eo^{\ar}_{l,\alpha},\eo^{\dr}_{l,\alpha})$ and $c((\vo,\emptyset))\geq c((\vo,(\eo \setminus (\eob^{\ar}_{l,\alpha}\cup \eo^{\ar}_{l,\alpha}) \cup \eo^{\dr}_{l,\alpha})))$ are always satisfied for any step $\alpha$.
   Hence, the attacker will choose to attack all edges with strong signals in any step $\alpha$ given enough energy. This can be achieved if the attacker has high enough stored energy, i.e., (\ref{lem1b1}) is satisfied, or if the attacker has high enough recharge rate, i.e., $\ra \geq \bab |\eo|$. These conditions enable the attacker to attack all edges strongly while still satisfying the energy constraint (\ref{a}) above for all steps. $\hfill \square$
    \end{pf}

    {\color{black}\begin{prop} \label{lem3.5}
    A sufficient condition for all agents not to achieve consensus at infinite time is that the attacker's parameters satisfy $\ra/\bab \geq |\eo|$.
    \end{prop}
    \begin{pf}
    By Lemma~\ref{lem1b}, the attacker always strongly attacks all edges with strong signals in a game at any step $\alpha$ given either sufficient recharge rate or sufficient stored energy at the beginning of the game. Consequently, if the attacker's recharge rate satisfies $\ra/\bab \geq |\eo|$, the attacker will attack $\eo$ with stronger jamming signals at all steps of all games, separating every agent at all times. As a result, there are $n$ clusters formed, and hence, obviously, consensus is not reached. $\hfill \square$
    \end{pf}}

    \begin{rem}
    
    {\color{black} Note that the necessary conditions and the sufficient condition above consider $z_k=x^{\mathrm{T}}L_{\mathrm{c}}x$ in (\ref{z}) which is a nonincreasing function. It is possible to consider other Laplacian matrices, e.g., Laplacian of the underlying graph $\go$, however the function $z_k$ may not be nonincreasing anymore. For example, we consider a simple path graph 1-2-3 with initial states $x_0=[10,0,-5]^{\mathrm{T}}$ and Laplacian of graph $\go$ considered in state difference function $z_k$. With weights of the utility functions (\ref{ua}) and (\ref{ud}) $a=1$ and $b=0$ and under consensus protocol (\ref{state}) and (\ref{state2}) with weights $a_{12}=0.1$ and $a_{23}=0.8$, the players' utilities in the first game with $h=1$ are $U^{\ar}_{1}=-U^{\dr}_{1}=148$ without any attacks, and $U^{\ar}_{1}=U^{\ar}_{0}=-U^{\dr}_{1}=125$ if both edges are attacked. This implies that not attacking any edge may actually be optimal for the attacker even with large enough energy. As a consequence, with Laplacian of graph $\go$ considered in state difference function $z_k$, the analysis becomes more complicated and some of the theoretical results do not hold anymore, e.g., the sufficient condition in Proposition~\ref{lem3.5}.}
    \end{rem}
    
    \subsection{\textcolor{black}{Example on a Gap Between Necessary Condition and Sufficient Condition}}
    
   {\color{black} In this subsection we provide an example that illustrates the gap between the necessary condition for preventing consensus in Theorem~\ref{lem3} and the sufficient condition in Proposition~\ref{lem3.5}. Here we suppose that the defender has a very high recharge rate (i.e., $\rd$ is much larger than $\bd$) such that it can recover any normally-attacked edges at any $k$ (note that the condition in Theorem~\ref{lem3} only consists of the attacker's parameters). This will force the attacker to attack with strong jamming signals to disconnect any agent.}

\begin{figure}[t]
        \centering
        \includegraphics[scale=0.4]{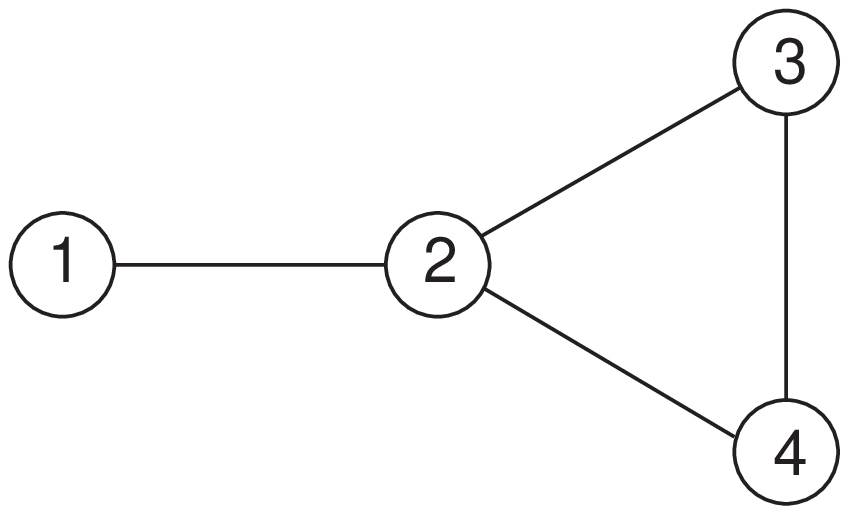}
        \caption{Graph $\go$ used in the case study.}
        \label{difhorb}
    \end{figure}

    {\color{black} We consider a graph $\go$ as in Fig. \ref{difhorb}, with $x[0]=[-5,0,-20,10]$, $h=2$, and $\ka=\ra$. The weight of the utility functions are set to be $a=1$ and $b=0$. We test various values of $1 \leq \ra/\bab \leq 2$, implying that the attacker can attack one edge with strong signals at all time without running out of energy. Thus, the attacker needs to attack $e_{12}$ (min-cut edge of $\go$) at all times in order to prevent consensus, since it is the only edge which, if attacked, will make the graph disconnected. Note that this ratio $1 \leq \ra/\bab \leq 2$ satisfies the necessary condition for preventing consensus in Theorem~\ref{lem3}, but not the sufficient condition in Proposition~\ref{lem3.5}.} 

{\color{black} Specifically in this example we test whether consensus is prevented or not for various value of $\ra/\bab$ based on agent states at time $k=20$. It is interesting to note from Table~\ref{exnew} that even with a relatively small value of $\ra/\bab<|\eo|$, consensus can still be prevented by the attacker. 

From this example, we observe that there is a gap between the necessary condition and the sufficient condition. Note that this gap may be larger for a more connected $\go$ as well as for network consisting of more agents, where typically $|\eo|>>\lambda.$ Later in Section~\ref{sec6}, we provide more detailed examples which illustrate the effect of these parameters' values on consensus.
}  
 \begin{table}
        \caption{Agent state difference for various values of $\ra/\bab$}
        \renewcommand{\arraystretch}{1.4}
        \label{exnew}
        \centering
        \begin{tabular}{|c|c|c|}
        \hline
           $\ra/\bab$ & $z_{20}$ & Consensus  \\
            \hline
            1 & $0.113$ & Yes \\
            \hline
            1.1 & $0.115$ & Yes \\
            \hline
            1.2 & $1.3405$ & No \\
            \hline
            1.4 & $235.345$ & No \\
            \hline
            1.8 & $706.8$ & No \\
            \hline
            2 & $1354$ & No \\
            \hline
            \end{tabular}
        \vskip -5pt
    \end{table}

{\color{black} As the last result of the section, we state that for a special case with the complete graph under $b=0$ and $h=1$, i.e., a single-step game without rolling horizon, the condition in Theorem~\ref{lem3} is also sufficient, i.e., there is no gap between the necessary condition and the sufficient condition.}
    
    \begin{prop} \label{lem3.2}
      Suppose that $b=0$ and $h=1$. In the complete graph $\go$, a sufficient condition for consensus not to happen is $\ra/\bab \geq n-1$.
    \end{prop}
    

    \begin{pf}
    With $h=1$, the attacker will spend all of its energy at the only step of the game. With $\ra/\bab\geq n-1$, the attacker is always able to disconnect the complete graph $\go$. 
    
    {\color{black} In the complete graph $\go$, every agent is connected to all other agents regardless of their states, implying that there is no agent that can be prioritized to be isolated by the attacker (different from the example above). Then, with $b=0$, the attacker is ensured to separate the furthest agent. This implies that, at each game (and at each $k$), the attacker will always attack the same edges, resulting in disconnected $\gd$ at each time. $\hfill \square$}
    \end{pf}
    
    {\color{black} We note that in different class of graphs (including in other symmetric graphs such as cycle graphs or star graphs), it is more challenging to derive a tighter sufficient condition. This is because agents have direct access only to some other agents which makes cluster forming based on the agent states more difficult.}
\vspace{1cm}
    
    \section{Clustering Analysis}\label{thrf}
    
    In this section, we derive some results on the number of formed clusters of agents at infinite time. From Proposition~\ref{lem3.5}, the result implies the simple case where if the attacker has enough energy such that $\ra/\bab \geq |\eo|$, then the attacker can attack all the edges of the underlying topology $\go$ so that the number of clusters is $n$ (i.e., all the agents are separated).

    The next result discusses a relation between the attacker's cost and energy recharge rate with the maximum number of clusters that the attacker may create through jamming. In the subsequent results of this section, we suppose that $b=0$.
    
     We first define a vector which characterizes the maximum number of clusters of $\go$, given the parameters $\ra$ and $\bab$. Specifically, we define a vector $\Theta\in \mathbb{R}^{|\eo|}$ with elements $\Theta_j:=\max_{|\eo^{\ar}|=j} \overline{n}(\vo,{\eo}\setminus{\eo^{\ar}})$, with $\overline{n}(\vo,{\eo}\setminus{\eo^{\ar}})$ being the number of agent groups of $(\vo,{\eo}\setminus{\eo^{\ar}})$.

    

    \begin{prop} \label{lem5.5}
        An upper bound on the number of formed clusters at infinite time is $
        \Theta_{\floor{\ra/\bab}}$.
    \end{prop}
    \begin{pf}
    The vector $\Theta$ consists of the maximum number of formed groups $\overline{n}(\vo,{\eo}\setminus{\eo^{\ar}})$ given the number of attacked edges as the element index. Since some edges need to be attacked consistently in order to divide the agents into different clusters, the number of formed clusters at infinite time is never more than the maximum number of groups at any time $k$ given the same number of strongly attacked edges.
    
    Recall that $\floor{\ra/\bab}$ is the maximum achievable number of edges that can be strongly attacked at all times. Given the known graph topology $\go$, we then can imply that $\Theta_{\floor{\ra/\bab}}$ gives the maximum number of clusters at infinite time.  $\hfill \square$
    \end{pf}
    \vspace{0.0cm}

    We continue by addressing a special case where all the agents in the network are connected with each other.
    
    \begin{cor} \label{lem6}
        {\color{black} In the complete graph $\go$, the attacker cannot divide the agents into more than 
        \begin{align}
        1+\sum_{j=1}^{(n-1)} \min\Bigl\{1,\Bigl\lfloor{\frac{2\ra}{j \bab(2n-j-1)}\Bigr\rfloor}\Bigr\} \label{compl}
        \end{align}
        number of clusters.}
    \end{cor}
    \begin{pf}
    {\color{black} In the complete graph, every agent is connected to all other $n-1$ agents. From Proposition~\ref{lem5.5}, we can derive the vector $\Theta$ of the complete graph $\go$ as 
    \begin{align*}
        \Theta=&[1,\ldots,1, 2,\ldots,2,3,\ldots,n-1,n]^{\mathrm{T}},
    \end{align*}
    where the value of the $(n-1)$th entry is 2, the value of the $((n-1)+(n-2))$th entry is 3, and so on. This is because in the complete graph $\go$ the attacker needs to attack $(n-1)$ number of edges to disconnect the graph, further $(n-2)$ number of edges to make three groups of agents, further $(n-3)$ number of edges to make four groups of agents, and so on, until $(n-1)+(n-2)+\cdots+1=n(n-1)/2$ agents to make $n$ groups. The value of the $\floor{\ra/\bab}$th entry of this $\Theta$ matrix for the complete graph can be written as in (15). This value determines the upper bound of the number of clusters.} $\hfill \square$
    \end{pf}
    \vspace{0.0cm}
    
    
    \blfootnote{In Proposition~\ref{lem5.5}, we use the information of the graph structure to obtain the vector $\Theta$. We remark that if the graph structure $\go$ is not known, then the number of clusters at infinite time is in general upper bounded by $\floor{{\ra/\bab}}+1$. This is because the attacker can attack continuously at all time at most $\floor{\ra/\bab}$ number of edges, and in the most vulnerable graph with $\lambda=1$, i.e., tree graphs, any attacked edge will result in a new group.
    
    To illustrate the relationship between $\Theta$ and $\ra/\bab$, we look at the graph in Fig.~\ref{difhorb} from the last section. Here,  $\Theta=[2,2,3,4]^{\mathrm{T}}$, whereas the values of $\floor{{\ra/\bab}}+1$ are 2, 3, 4, 5 for $\ra/\bab=1$, 2, 3, and 4, respectively. Note that for any value of $\ra/\bab$, inequality $\Theta_{\floor{\ra/\bab}}\leq \floor{\ra/\bab}+1$ is always satisfied, indicating that knowing the graph structure helps to better estimate the upper bound of the number of clusters.}

\section{Equilibrium Characterization}\label{thre2}
\suppressfloats
    In this game the strategy choices are all finite in form of edges attacked and recovered. Here, we characterize the equilibrium/optimal strategies of the players in certain situations for the case where the players' horizon length is 1 so that they myopically update their strategies every time step. 
   
    In this section, we state some results when $a=0$, i.e., when the players do not consider the agents' states but agent-group index in determining their strategies so that the defender (resp., attacker) has higher (resp., lower) utility when more agents belong to the same group. Similar to the analysis in \cite{arxivYur}, here we explore some possible optimal strategy candidates for the players in a game. However, since a game consists of several steps in this formulation, the subgame perfect equilibrium is more involved to characterize, compared to the case of a game consisting of one step as in \cite{arxivYur}.
    
    In the $\lb$th step of each game, there are three possibilities in function $c(\cdot)$ as shown in Table~\ref{tab:my_label} (Cases~1,~2,~and~3). From this table, we characterize the optimal strategies of both players in each case:

    \begin{table}
        \caption{Possible cases of attack and recovery actions}
        \renewcommand{\arraystretch}{1.4}
        \label{tab:my_label}
        \centering
        \begin{tabular}{|c|c|c|}
        \hline
            Case & $\ca$ & $\cd$  \\
            \hline
            1 & $\ca=\co$ & $\cd=\ca$ \\
            \hline
            2 & $\ca < \co$ & $\cd = \ca$ \\
            \hline
            3 & $\ca < \co$ & $\cd > \ca$ \\
            \hline
            \end{tabular}
        \vskip -5pt
    \end{table}
    
    \begin{itemize}
        \item \textbf{Case~1:} When $\co=\cd$, the attacker's utility in one time step is $\co$, which implies that the attacker should not attack any edge either with normal signals or strong signals, with the utilities of both players equal to zero. The players' strategies in this case are called Combined Strategy~1. 
        \item \textbf{Case~2:} When $\cd=\ca$, the defender does not recover any attacked edge, whereas the attacker should attack some edges either with strong or normal signals. The players' strategies in this case are classified as Combined Strategy~2. 
        \item \textbf{Case~3:} Here both players will attack/recover nonzero number of edges. In particular, the attacker will attack with normal signals and potentially with strong signals. The players' strategies here are called Combined Strategy~3. 
    \end{itemize}
     
    We will then discuss the equilibrium for this game in Proposition~\ref{lem03} below. For simplicity, we only consider the case when $h=1$. The case of $h>1$ can be examined based on the characterization here for $h=1$.

    \begin{prop} \label{lem03}
        The optimal strategies for the players with $h=1$ satisfy the following:
        \begin{enumerate}
            \item Combined Strategy~1 if $\tilde{\beta}^{\ar}_{l}+\ba>\ka + \ra (l-1)T$,
            \item Otherwise, 
            \begin{enumerate}
                \item Combined Strategy~2 if
                \begin{enumerate}
                    \item $\tilde{\beta}^{\dr}_{l}+\bd> \kd + \rd (l-1)T$, or 
                    \item $\tilde{\beta}^{\dr}_{l}+\bd\leq \kd + \rd (l-1)T$ and $\ua(\floor{(\ka + \ra (l-1)T-\tilde{\beta}^{\ar}_{l})/ \bab},\emptyset,\emptyset)=\max_{\eab,\ea,\ed} \ua(\eab,\ea,\ed)$,
                \end{enumerate}
                \item Combined Strategy~3 if $\tilde{\beta}^{\dr}_{l}+\bd\leq \kd + \rd (l-1)T$ and $\ua(\floor{(\ka + \ra (l-1)T-\tilde{\beta}^{\ar}_{l})/ \bab},\emptyset,\emptyset)\ne\max_{\eab,\ea,\ed} \ua(\eab,\ea,\ed)$.
            \end{enumerate}
        \end{enumerate}
    \end{prop}
    
    \begin{pf}
    With $a=0$, we observe that the defender always recovers from the optimal attack at the last step given sufficient energy, which implies that it always recovers for $h=1$ if $\tilde{\beta}^{\dr}_{l}+\bd\leq \kd + \rd ((l-1)T)$ is satisfied. Similar to the defender, the attacker obtains the least utility, i.e., zero, by not attacking for the case of $h=1$. Therefore, the attacker will attack at least one edge as long as it has enough energy to do so. We prove each point of the proposition statement as below.
    
    \textbf{(1):} We now suppose that $\tilde{\beta}^{\ar}_{l}+\ba>\ka + \ra ((l-1)T)$ (point (1) in the statement) is satisfied, i.e., the attacker does not have enough energy to even attack one edge normally. In this case, Combined Strategy~1 becomes optimal since there is no other choice, i.e., the attacker cannot attack even one edge with normal signals. In the rest of the proof, we assume that $\tilde{\beta}^{\ar}_{l}+\ba\leq \ka + \ra ((l-1)T)$ is satisfied.
    
    \textbf{\boldmath $(2a(i))$:} We now continue by providing the conditions for Combined~Strategy~2. Similarly to the attacker above, we observe that the defender cannot recover any edge if  $\tilde{\beta}^{\dr}_{l}+\bd > \kd + \rd ((l-1)T)$, implying that $\ca<\co$ and $\cd=\ca$ (corresponds to point $(2a(i))$). 
    
    \textbf{\boldmath $(2a(ii))$:} We then suppose that $\tilde{\beta}^{\dr}_{l}+\bd\leq \kd + \rd ((l-1)T)$ is satisfied. It then follows that given enough energy for the defender, the attacker needs to attack nonzero number of edges with strong signals to satisfy $\ca<\co$ and $\cd=\ca$. In order for Combined Strategy~2 to be optimal, the attacker then needs to attack edges strongly without attacking with normal signals at all, i.e., $\ea=\emptyset$. Thus, $\bab$ needs to be sufficiently low to make strong attack feasible. Specifically, $\ua(\eob^{\ar\prime}_k,\emptyset,\emptyset)=\max_{\eab,\ea,\ed} \ua(\eab,\ea,\ed)$, with $|\eob^{\ar\prime}_k|=\floor{(\ka + \ra ((l-1)T)-\tilde{\beta}^{\ar}_{l})/ \bab}$ indicating the maximum number of edges the attacker attacks strongly. This corresponds to point $(2a(ii))$.
    
    \textbf{\boldmath $(2b)$:} Consequently, if $\tilde{\beta}^{\dr}_{l}+\bd\leq \kd + \rd ((l-1)T)$ and $\ua(\eob^{\ar\prime}_k,\emptyset,\emptyset)\ne\max_{\eab,\ea,\ed} \ua(\eab,\ea,\ed)$ are true, then the attacker normally attacks nonzero number of edges and the defender recovers nonzero number of edges, which imply that Combined Strategy~3 is optimal (point $2b$). $\hfill \square$
    \end{pf}
    
    \begin{rem}{\color{black} The characterization of optimal strategies in Proposition~\ref{lem03} also holds for a more general class of agent-group indices other than $c(\go^{\prime})$ defined in (\ref{cluster}), as long as the utility function structure (\ref{ua}) and (\ref{ud}) does not change. Specifically, it holds for those indices that belong to the class given by
    \begin{align}
        \mathcal{C} := \{& \tilde{c} : 2^{\vo} \times 2^{\eo} \rightarrow \mathbb{R} : \tilde{c}((\vo,\overline{\eo}\cup\eo^{\prime}))\geq  \tilde{c}((\vo,\overline{\eo})),  \notag \\ &
          \overline{\eo},\eo^{\prime} \subseteq \eo\}. \label{asum2}
    \end{align}
    The condition $\tilde{c}((\vo,\overline{\eo}\cup\eo^{\prime}))\geq  \tilde{c}((\vo,\overline{\eo}))$ implies that not attacking results in the maximum value of $\tilde{c}(\go^{\ar}_{l,\lb})$ of the attacker. Similarly, for the defender, this condition implies that not recovering given the attacks results in the minimum value of $\tilde{c}(\go^{\dr}_{l,\lb})$. This condition is necessary for ensuring the equilibrium as in Proposition~\ref{lem03}, since it guarantees that attacking/recovering nonzero number of edges (corresponding to Combined~Strategy~3) is always optimal for the players as long as they have the energy to do so.}
    \end{rem}

    In general, since the cases discussed above are for one step only, for longer $h>1$ the optimal strategies will take form of a set of combined strategies. For example, if $h=3$, the sequence of optimal strategies may be \{Combined Strategy 1, Combined Strategy 2, Combined Strategy 2\}. On the other hand, for $a>0$, the condition in Proposition~\ref{lem03} becomes more complicated to characterize since attacking more edges does not necessarily result in the highest possible utility.

\section{Simulation Results}\label{sec6}
    \begin{figure}[t]
    \hskip -10pt
    \includegraphics[trim=8 0 0 0,clip, scale=0.6]{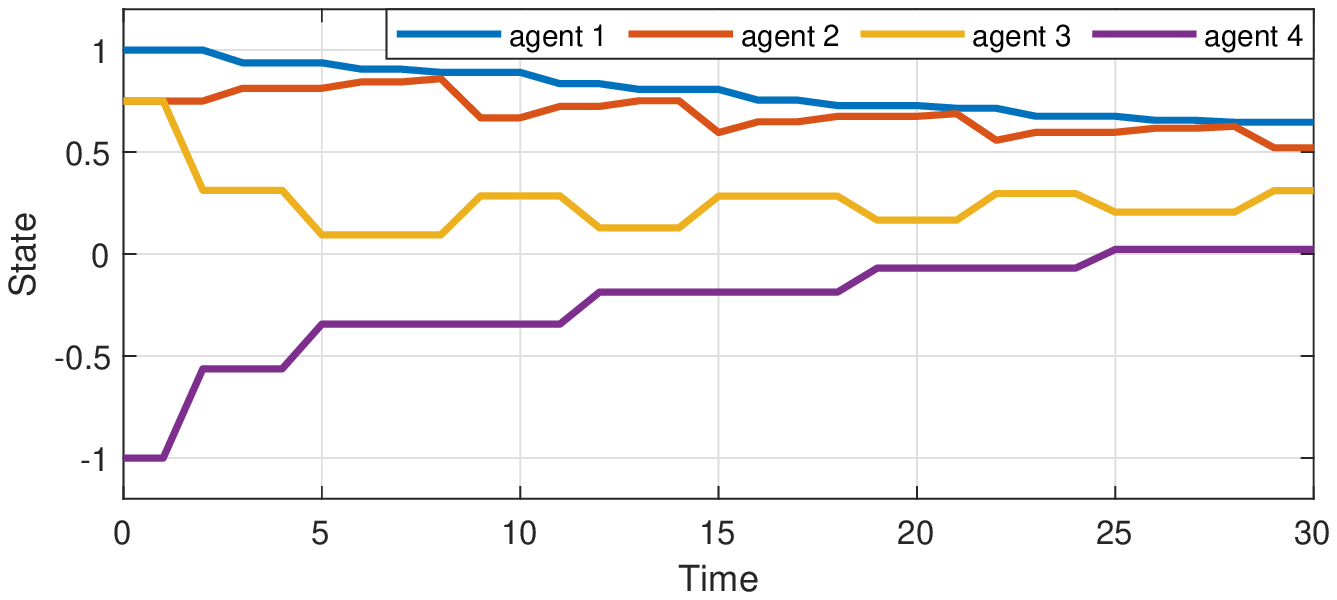}
    \vskip -7pt
    \caption{\small \textcolor{black}{Agent states with $a=0.1$ and $b=0.9$}}\label{fig1}
    \vskip 5pt
    \hskip -10pt
    \includegraphics[trim=8 0 0 0,clip, scale=0.6]{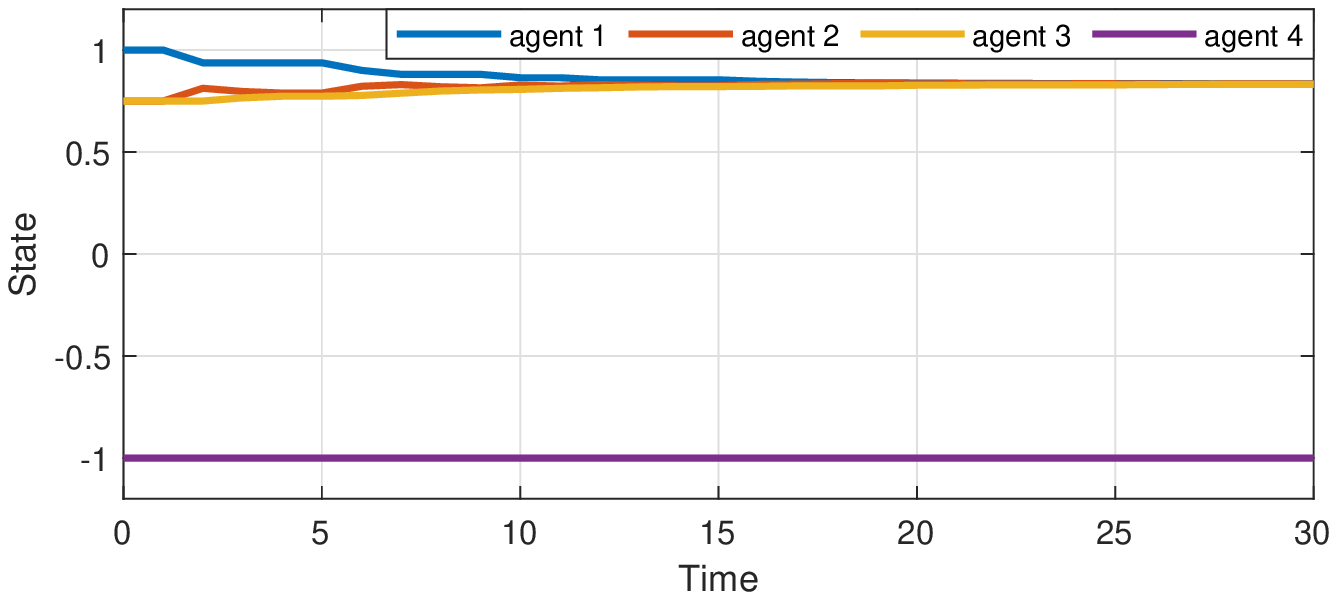}
    \vskip -7pt
    \caption{\small \textcolor{black}{Agent states with $a=0.9$ and $b=0.1$}}
    \label{fig3}
    \vskip 5pt
    \hskip -10pt
    \includegraphics[trim=8 0 0 0,clip, scale=0.6]{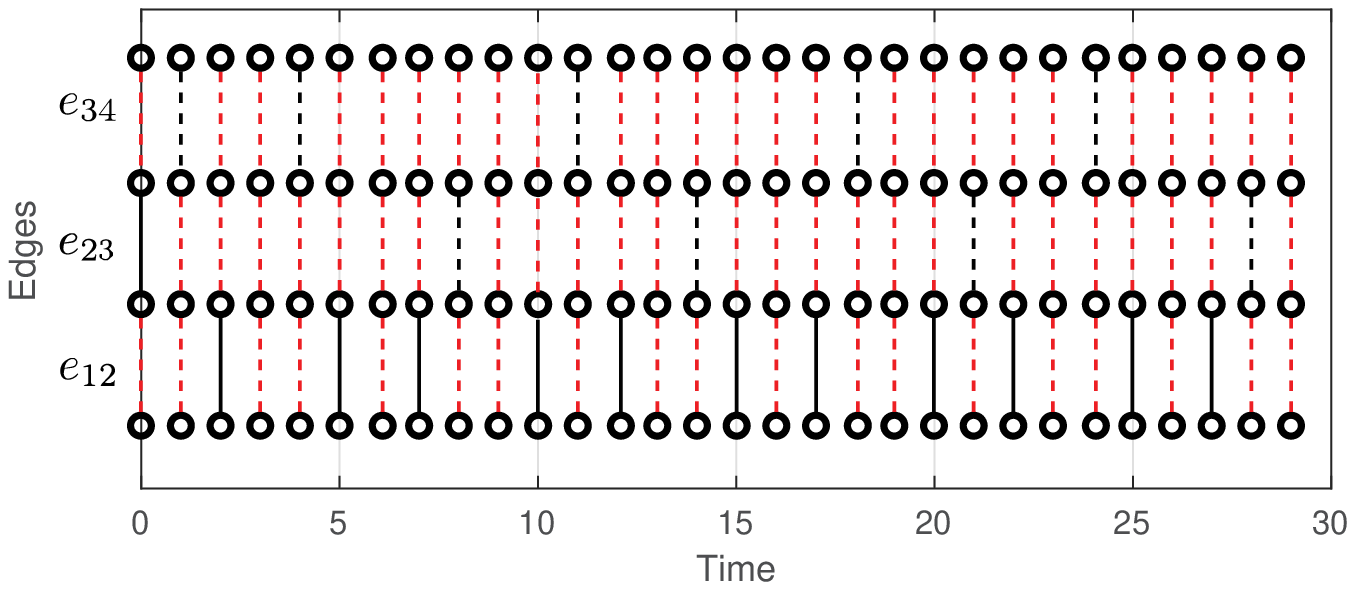}
    \vskip -7pt
    \caption{\small \textcolor{black}{Attacked and recovered edges with $a=0.1$ and $b=0.9$}}
    \label{fig2}
    \vskip 5pt
    \hskip -10pt
    \includegraphics[trim=8 0 0 0,clip, scale=0.6]{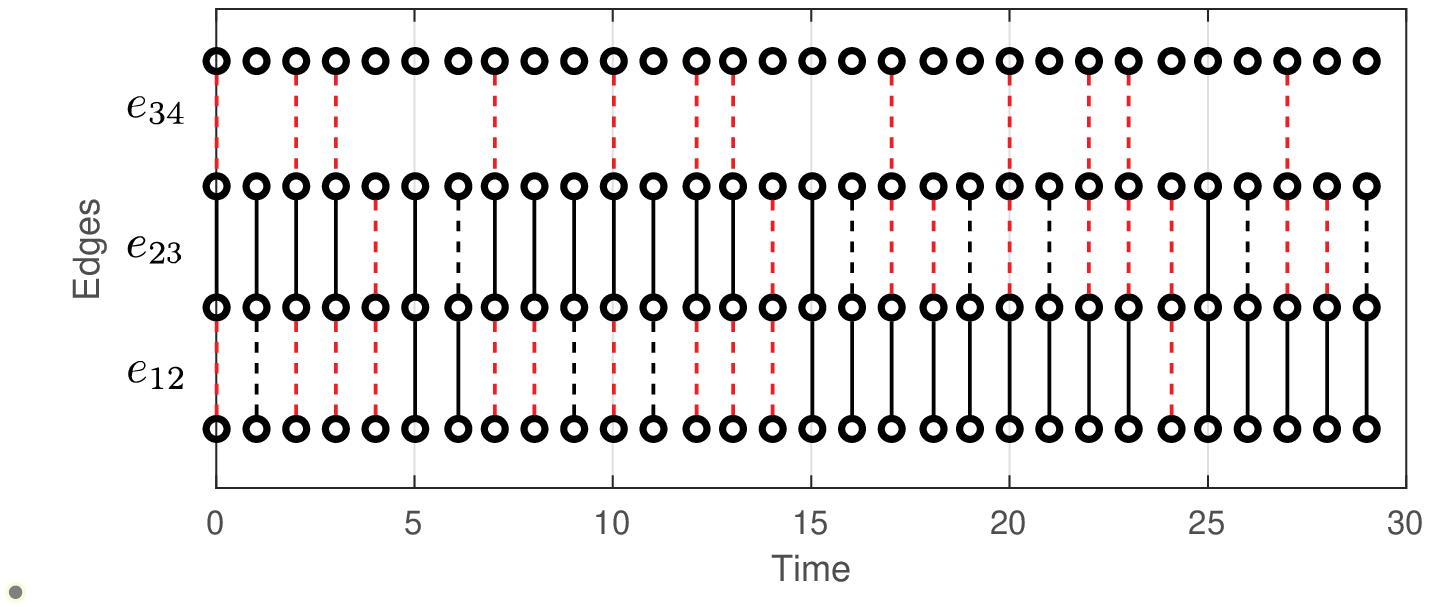}
    \vskip -7pt
    \caption{\small \textcolor{black}{Attacked and recovered edges with $a=0.9$ and $b=0.1$}}
    \label{fig4}
\end{figure}

\subsection{Consensus and Clustering across Parameters}\label{satH}

Here we show how the consensus varies across different weights of the utility functions and the initial states.

\subsubsection{Varying Weights $a$ and $b$}\label{ab}

We consider the 4-agents line/path graph $1$--$2$--$3$--$4$ with initial states $x_0=[1,0.75,0.75,-1]^{\mathrm{T}}$. {\color{black} The parameters are $\ba=\bd=1$, $h=\bab=2$, $\ka=\ra=2.6$, $\rd=0.3$, and $\kd=0.8$, which satisfy the necessary condition for preventing consensus in Proposition~\ref{lem3.4}, but not the sufficient condition in Proposition~\ref{lem3.5}}. With $b=1-a$, Figs.~\ref{fig1} and \ref{fig3} show the agent states with small $a$ (at $a=0.1$) and large $a$ (at $a=0.9$), respectively. Figs.~\ref{fig2}~and~\ref{fig4} illustrate the status of the edges in $\gd$ over discrete time $k$. There, no line in the corresponding edge implies that the edge is strongly attacked; likewise, dashed red lines: normally attacked, dashed black lines: recovered, and solid black lines: not attacked. 

We observe that for small $a$, the attacker more often divides the agents into more groups, indicated by more dashed red lines in Fig.~\ref{fig2}. As a result, the attacker fails to prevent consensus among the agents (Fig.~\ref{fig1}), despite the condition in Proposition~\ref{lem3.4} being satisfied. On the other hand, with large $a$, the attacker is more focused to make the difference among agents' states larger while separating the agents into fewer groups compared to the case with small $a$. These features can be seen in Fig.~\ref{fig4}, where there are no black lines in the edge $e_{34}$, and thus no consensus among the agents in Fig.~\ref{fig3}. 

\begin{figure}
    \hskip -15pt
    \includegraphics[trim=10 0 0 0,clip, scale=0.6]{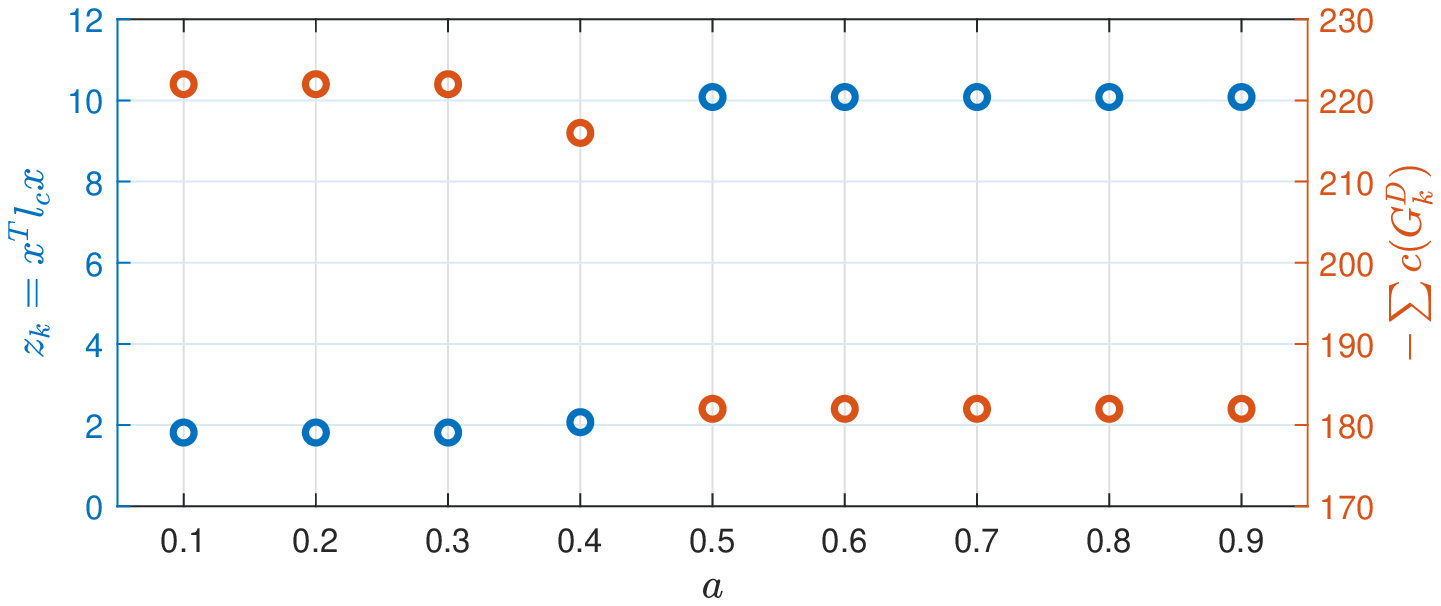}
    \vskip -10pt
    \caption{\small Comparison of $z_k$ and $- \sum c(\gd)$ ($k=20$) versus $a$}
    \label{fig5}
    \vspace{0.5cm}
    \centering
    \includegraphics[scale=0.6]{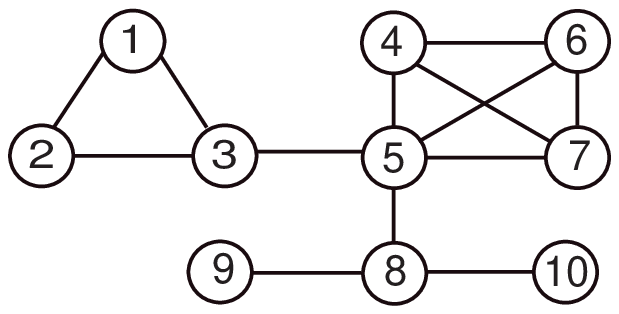}
    \caption{Graph used for simulation in Section~\ref{satF}}
    \label{fig16}
    \vskip 0pt
\end{figure}

We next present a comparison in the optimal state difference $z_k(\eob^{\mathrm{A}*}_k,\eo^{\mathrm{A}*}_k,\eo^{\dr*}_k)$ and agent-group index $c(\gd)$ across different $a$ and $b=1-a$ in Fig. \ref{fig5}. We observe that with larger $a$, the attacker successfully prevents consensus among agents (shown with larger value of $z_k$) at time $k=20$. On the other hand, with smaller $a$ (corresponding to larger $b$), the attacker obtains higher $c(\go^\dr_k)$ at the cost of low $z_k$, implying that the attacker fails to prevent consensus. It is interesting that the values of $z_k$ and $\sum c(\go^\dr_k)$ remain almost constant for some different $a$, implying that there is a critical value of weights $a$ and $b$ that determine the consensus and the number of clusters at infinite time; in this case, the critical value of $a$ is located in $0.4<a<0.5$.

\subsubsection{Varying Initial States $x_0$}\label{satF}

\begin{figure}
    \hskip -10pt
    \includegraphics[trim=8 0 0 0,clip, scale=0.6]{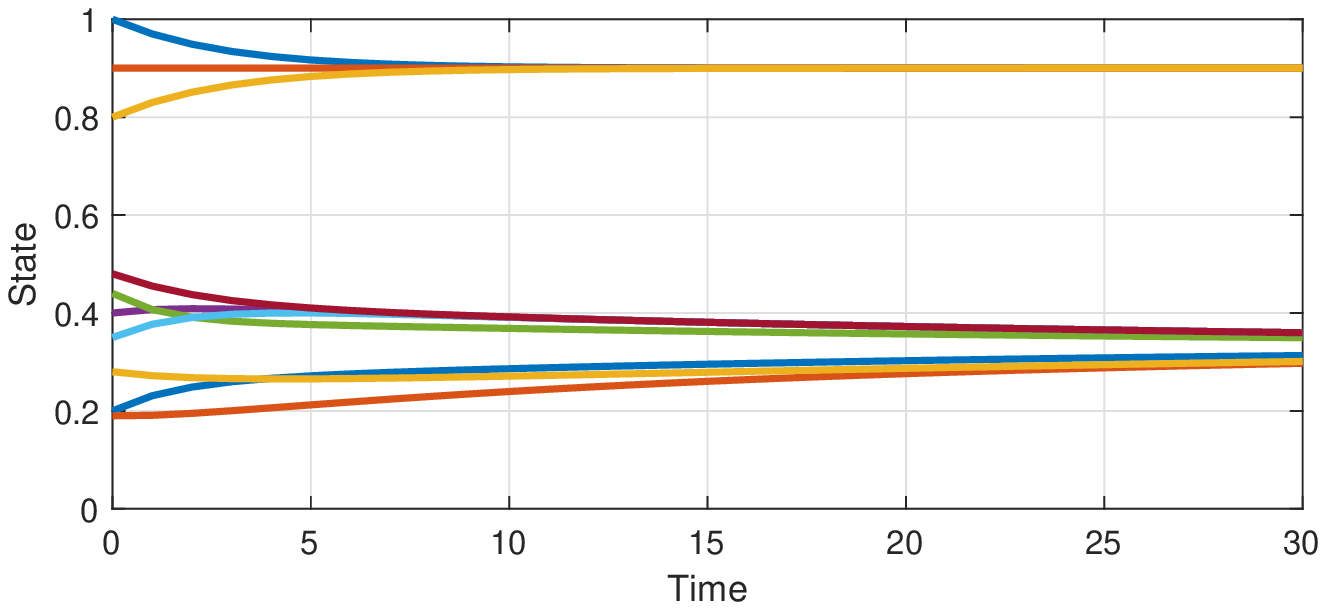}
    \vskip -10pt
    \caption{\small Agent states in Case~1}
    \label{fig7}
    \vskip 5pt
    \hskip -10pt
    \includegraphics[trim=8 0 0 0,clip, scale=0.6]{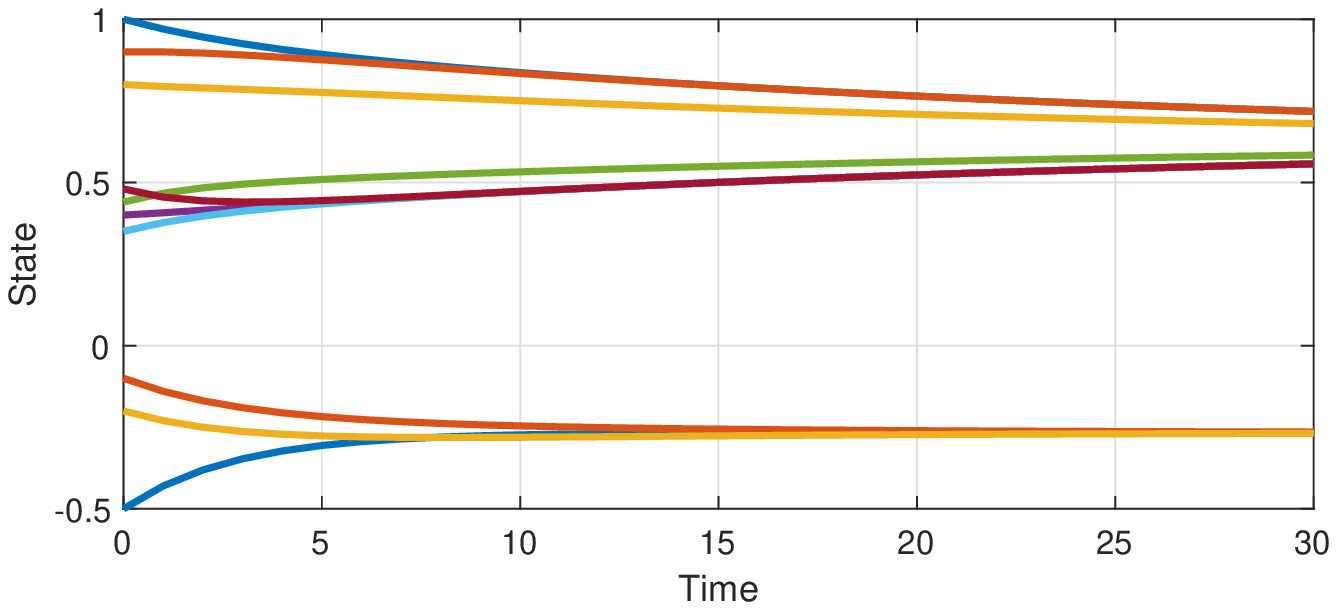}
    \vskip -10pt
    \caption{\small Agent states in Case~2}
    \label{fig8}
    \vskip 5pt
    \hskip -10pt
    \includegraphics[trim=8 0 0 0,clip, scale=0.6]{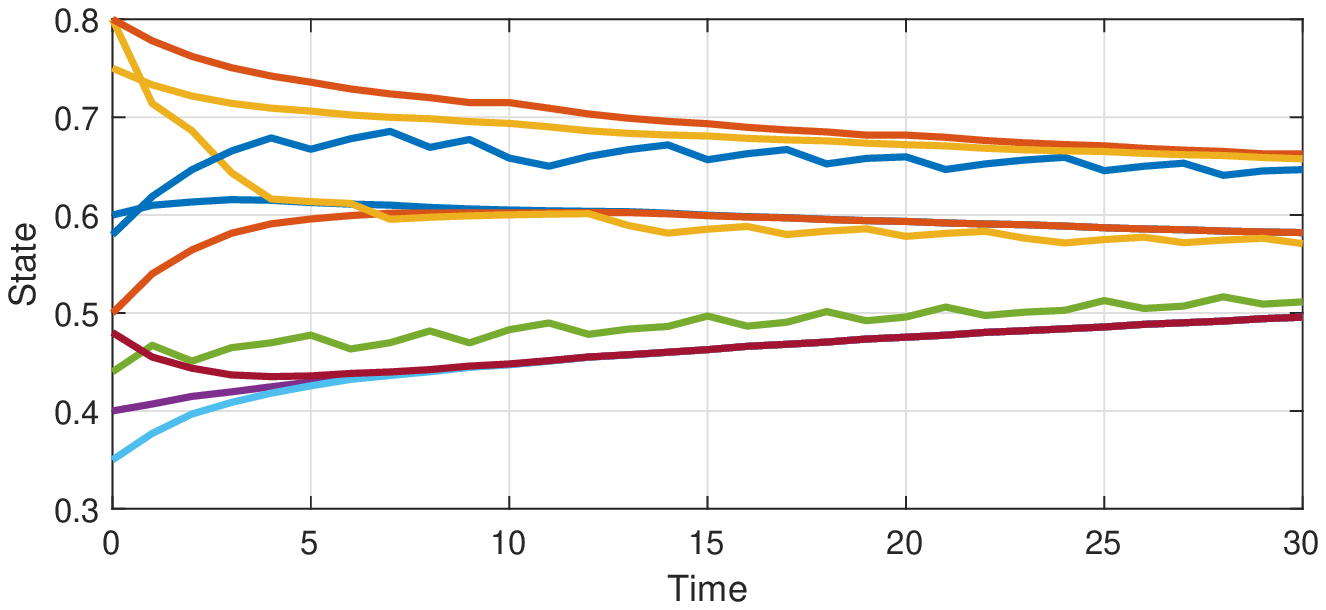}
    \vskip -10pt
    \caption{\small Agent states in Case~3}
    \label{fig9}
\end{figure}

We also observe how the initial states $x_0$ affect the agent-group index of the agents. We consider the graph shown in Fig.~\ref{fig16}, which consists of 10 agents. All parameters other than the initial states are set to be the same and satisfy the conditions in Proposition~\ref{lem3.4}. Specifically, we set $\ba=\bd=1, \bab=2, \ka=\ra=2.1,\kd=\rd=0.7$, and $a=1-b=0.9$. The state trajectories of the agents with varying $x_0$ are shown in Figs.~\ref{fig7}--\ref{fig9}. Here we consider three cases of initial states $x_0$:
\begin{enumerate}
    \item \small $x_0=[1,0.9,0.8,0.4,0.44,0.35,0.48,0.2,0.19,$ $0.28]^{\mathrm{T}}$,
    \item \small $x_0=[1,0.9,0.8,0.4,0.44,0.35,0.48,-0.5,-0.1,-0.2]^{\mathrm{T}}$,
    \item \small $x_0=[0.6,0.5,0.8,0.4,0.44,0.35,0.48,0.58,0.8,0.75]^{\mathrm{T}}$.
\end{enumerate}
 Note that in Case~(1), agents 1--3 have closer initial states and are far from the other agents. Similarly, in Case~(2), agents 8--10 have initial states that are different from the other agents. However, in Case~(3), agent states are distributed approximately evenly in the range $[0.35,0.8]$ so that it is hard for the attacker to divide them into clusters.

From Fig.~\ref{fig7}, we can see that in Case~(1), agents~1--3, which have weak connection to other agents (only connected by one edge), are grouped together and converge to the same state. This occurs by attacking the edge connecting agents 3 and 5. On the other hand, in Fig.~\ref{fig8} for Case~(2), agents 8--10 are separated from the others because the edge connecting agents 5 and 8 is attacked continuously. Clearly, in Cases~(1) and (2) it is easier for the attacker to separate agents since their initial states form clusters matching
the network topology.

In Case~(3), however, the initial state values do not exhibit such properties and
as a result, the states converge towards the same value as shown in Fig.~\ref{fig9}. In this simulation, the attacker is not able to effectively attack certain edges at all times; as a consequence, the agents are not divided into clusters and thus consensus happens. The attacker may be able to prevent consensus with higher weight $a$, as discussed in Section~\ref{ab} above.

For obtaining Figs.~\ref{fig7}--\ref{fig9}, we solve combinatorial optimization problems to find optimal strategies of the players. We remark that the computational complexity of this problem depends on the number of edges $\eo$ of $\go$. We have reduced the complexity by disregarding some combinations of edges that are clearly not optimal; for example, attacking only the edge connecting agents 4 and 7 does not disconnect the graph, and thus cannot be the best move for the attacker.

\subsubsection{Varying Energy and Cost Parameters}\label{satG}
We continue by discussing the effect of the attacker's recharge rate $\ra$ and unit costs of attacks $\ba$ and $\bab$ on the consensus and cluster forming. Recall that in the theoretical results in Sections~\ref{thre} and \ref{thrf}, the ratios of $\ra$ to $\bab$ and $\ra$ to $\ba$ are used to derive the necessary conditions and sufficient conditions for preventing consensus as well as the upper bound of the number of clusters formed at infinite time.

\begin{figure}
    \hskip -10pt
    \psfrag{ed6}{$|\eo|$}
    \psfrag{ed7}{$\frac{\ra}{\bab}$}
    \includegraphics[trim=0 0 0 0,clip, scale=0.6]{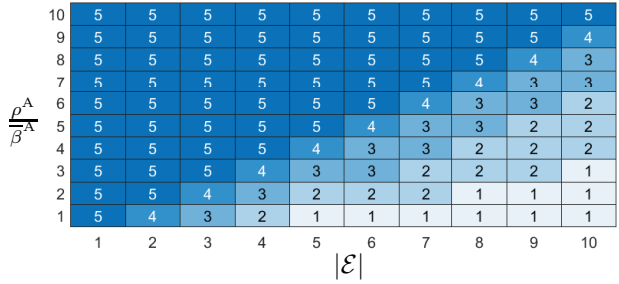}
    \vskip -5pt
    \caption{Number of clusters at $k=50$ with $b=0$. The underlying graphs used are those with $5$ agents with maximum $10$ edges.}
    \label{fig13}
\end{figure}

Assuming that $b=0$, the number of clusters is dictated by $\ra/\bab$ as discussed in  Proposition~\ref{lem5.5}. We show the number of clusters over different topologies of the underlying graph $\go$ in Fig.~\ref{fig13}. We consider networks with $n=5$, with the edges positioned to yield the most connected topology, i.e., maximum $\lambda$, given the same number of edges $|\eo|$. Note that, with $n=5$, there are at most $n(n-1)/2=10$ number of edges in the underlying graph $\go$ (which happens for the complete graph $\go$). We observe that with $\ra/\bab\geq|\eo|$, the agents are divided into 5 clusters (all agents are separated) as shown in the upper left area of the figure indicated by \say{5} as derived in Proposition~\ref{lem3.5} whereas in the lower right area indicated by \say{1} the agents converge to the same cluster. It is clear that in a more connected graph, the agents are more likely to converge to a fewer number of clusters.

\subsection{Players' Performance Under Varying Horizon Length and Game Period}
In this subsection, we evaluate the players' performance under varying horizon length $h$ and game period $T$. To evaluate the performance of the players, we introduce the \textit{applied utilities} $\uah:=a z_k(\eob^{\ar*}_k,\eo^{\ar*}_k,\eo^{\dr*}_k) - bc(\gdo)$ and $\udh:=-a z_k(\eob^{\ar*}_k,\eo^{\ar*}_k,\eo^{\dr*}_k) +b c(\gdo)$, with $\gdo=(\vo,((\eo\setminus(\eob^{\ar*}_k\cup \eo^{\ar*}_k))\cup\eo^{\dr*}_k)$. These are elements of utility functions $\ua$ and $\ud$ corresponding to the $\lb$th step, $\lb=k \ \mathrm{mod} \ {T}+1$, of the game with index $l=\floor{k/T}+1$, where the obtained strategies $(\eob^{\ar*}_{(l-1)T+\lb-1},\eo^{\ar*}_{(l-1)T+\lb-1},\eo^{\dr*}_{(l-1)T+\lb-1})=(\eob^{\ar*}_{l,\lb},\eo^{\ar*}_{l,\lb},\eo^{\dr*}_{l,\lb})$ are applied. Since $\ua=-\ud$, having higher applied utility for the attacker implies lower applied utility for the defender. Note that the values of $h$ and $T$ are uniform among the players.

In this subsection, we consider the weight $a_{ij}=\hat{a}$, $\hat{a}<1/n$ in (\ref{state2}) which implies that different agents have different convergence speeds depending on the number of their neighbors.
Furthermore, we consider various initial states $x_0$ for the agents in order to more accurately evaluate the attacker's performance and the pattern of applied utilities $\uah$. We use up to 1000 randomly generated initial states in this simulation  for each agent ranging from $-1$ to $1$. Throughout this subsection, we use parameters $n=3$, $\ra=1.1$, $\ka=7$, $\bab=2\ba=1$.

\subsubsection{Players' Performance Under Varying Horizon Length}
\begin{figure}
    \hskip -10pt
    \includegraphics[trim=8 0 0 0,clip, scale=0.6]{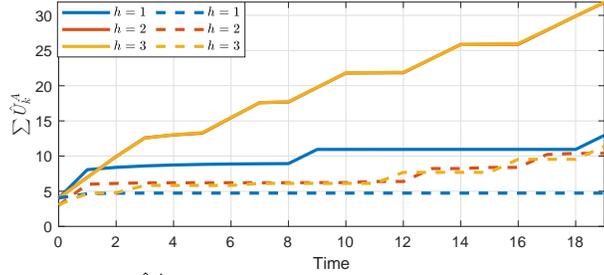}
    \vskip -10pt
    \caption{\small $\sum_k \uah$ in the path graph (solid lines) and the complete graph (dashed lines) for varying value of $h$. The applied utility for $h=2$ and $h=3$ in the path graph is almost identical.}
    \label{fig11a}
\end{figure}

\begin{table}[]
\vspace{10pt}
\caption{Difference in the optimal actions and the resulting utilities in the path graph $\go$ between $h=2$ and $h=3$}
\label{tab2}
\centering
\resizebox{8.5cm}{!}{\begin{tabular}{|l|c|c|c|c|}
\hline
\multirow{2}{*}{Initial states} & \multicolumn{2}{c|}{$|\eob^{\ar*}_0|$} & \multicolumn{2}{c|}{$\sum_{k=0}^{19} \uah$} \\ \cline{2-5}
 & $h=2$ & $h=3$ & $h=2$ & $h=3$ \\\hline
 \footnotesize $[0.824, \ -0.798,	\ -0.413]^{\mathrm{T}}$ & {2} & 2 & \multicolumn{2}{|c|}{37.74} \\ \hline  
 \footnotesize $[-0.983, \	0.649, \	0.535]^{\mathrm{T}}$ & 2 & 2 & \multicolumn{2}{|c|}{39.89}\\ \hline  
  $[-0.787,	\ -0.786,	\ -0.265]^{\mathrm{T}}$              & 2 & 1  & 28.41 & 30.00  \\ \hline
  $[0.624, \ 0.629,	\ -0.821]^{\mathrm{T}}$ &  2 & 1    & 37.92 & 43.45\\ \hline
\end{tabular}}
\end{table}


\begin{figure}
    \hskip -10pt
    \includegraphics[trim=8 0 0 0,clip, scale=0.6]{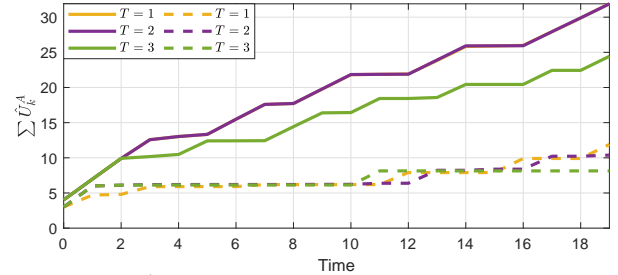}
    \vskip -10pt
    \caption{\small $\sum_k \uah$ in the path graph (solid lines) and the complete graph (dashed lines) for varying $T$. The applied utility for $T=1$ and $T=2$ in the path graph is almost identical.}
    \label{fig13a}
\end{figure}

We now consider the case of varying value of horizon length $h$ when the network is a path graph and a complete graph. Note that the value of $h$ is still uniform among the attacker and the defender. The evolutions of the attacker's applied utility $\uah$ with varying $h$ (with $T=1$ for every $h$) are shown in Fig.~\ref{fig11a}.

Since the path graph is the least connected graph, the attacker will be able to make multiple groups of agents relatively easily compared to more connected graphs. As a result, the attacker may not need to have a very long horizon length $h$ to improve its utility since it does not need to save energy as much compared to the case of the complete graph. This is shown with the overlapping red and yellow solid lines in the Fig.~\ref{fig11a}, implying that the horizon length $h=3$ is already as good as the case of $h=2$. On the other hand, the blue solid line is far below the red and the yellow ones, implying that having $h$ being too short can result in a worse utility for the attacker over time.

The differences of the attacker's strategies for some notable cases in the path graph $\go$ between $h=2$ and $h=3$ are shown in Table~\ref{tab2}. Here, we see the difference in the optimal actions between the attacker with $h=2$ and $h=3$ in the path graph $\go$ even though the plots of applied utilities in Fig.~\ref{fig11a} are very similar. We observe that when the initial states of some agents are sufficiently close, the attacker with $h=2$ keeps attacking both edges at $k=0$, whereas the attacker with $h=3$ chooses to save its energy by attacking fewer edges. At $k=19$ the attacker with $h=3$ obtains higher applied utility, indicating that it is able to better use its energy than the attacker with $h=2$ by attacking later.

On the other hand, since the complete graph is the most connected graph, here the attacker will need more energy to disconnect the graph and obtain some utility. Consequently, even with longer $h$, the difference of $\sum \uah$ is smaller compared to the path graph case. The difference between the red and the yellow dashed lines is clearer however, suggesting that the attacker still benefits by having $h=3$ (compared to the very little difference in the path graph case). The attacker's different behavior for the path graph and the complete graph $\go$ suggests that in a less connected graph, the effectiveness of longer $h$ may saturate from a lower value compared to the one in a more connected graph $\go$, given the attacker's energy parameters. 

In general, we observe that having a longer $h$ may result in a better applied utility for the attacker over time due to its role as a leader of the game, i.e., the attacker moves first and is able to choose its strategy that minimizes the defender's best response. Additionally, there is also a clear pattern on when $\sum \uah$ increases; this implies that the variation of initial states may not affect the attacker's optimal strategy, except in some cases as explained above.

{\color{black} We also remark that the effect of different values of $h$ is also influenced by the underlying graph $\go$. Specifically, in a less connected graph $\go$, having a very short horizon may even be more harmful compared to the case with a more connected $\go$. For example, in Fig.~\ref{fig13a}, the difference of $\sum \uah$ in the path graph between $h=1$ and $h=2$ is much more apparent than in the complete graph. The possible reason is that in the path graph, it is easier for the attacker to disconnect all agents and make $n$ groups at some time steps. Thus, with large enough $h$, the attacker can save enough energy to make $n$ groups more often. On the other hand, we also observe that increasing horizon length from $h=2$ to $h=3$ has minimal effect on the attacker's utility for the path graph, indicating that increasing horizon length past a certain value may not be beneficial anymore. As we see later, the similar phenomenon also happens for varying values of $T$.}

\subsubsection{Players' Performance Under Varying Game Period}

We then continue by simulating the case of varying value of game period $T$ (value of $h$ is set to be $h=3$ for both players so that the assumption $T\leq h$ is always satisfied). The average value of $\sum \uah$ over time is shown in Fig.~\ref{fig13a}, where in general, the attacker with shorter game period $T$ has higher applied utility especially at later time for both the path graph and the complete graph $\go$. 

The attacker with shorter $T$ will be more adaptive to the changes of the agents' and players' conditions. In the context of this game, the attacker with shorter $T$ may delay the attack further to maximize its utility later. This in turn increases the attacker's utility at later time, similar to the case of longer $h$ discussed above. Note that the yellow dashed and solid lines are the same as the yellow lines in Fig.~\ref{fig11a}, and we observe that the green and the purple lines do not differ as much as the red and the blue lines in Fig.~\ref{fig11a}, indicating that for the attacker, having a large value of $T$ may not be as disadvantageous as having short $h$.

\begin{table}[]
\vspace{10pt}
\caption{Average total number of edges attacked in the path graph $\go$}
\label{tab3}
\centering
\begin{tabular}{|l|c|c|c|c|c|}
\hline
\multirow{2}{*}{$h$} & \multirow{2}{*}{$T$} & \multicolumn{2}{c|}{$\sum_{m=0}^k|\eo^{\ar*}_m|$ (Normal) } & \multicolumn{2}{c|}{$\sum_{m=0}^k|\eob^{\ar*}_m|$ (Strong) } \\ \cline{3-6}
 &  & $k=9$ & $k=19$ & $k=9$ & $k=19$\\\hline
1    & \multirow{3}{*}{$1$} & 7 & 16 & 5 & 6 \\ \cline{1-1} \cline{3-6}
 2  &  & 0 & 0 & 8 &  13.959 \\ \cline{1-1} \cline{3-6}
 \multirow{3}{*}{$3$}    &  & 0 & 0 & 7.993 & 13.971    \\ \cline{2-6}
     & 2  & 0 & 0 & 8 & 13.970     \\ \cline{2-6}
  & 3 & 2.970 & 4.970 & 7.003 &    11.015  \\ \hline
\end{tabular}
\end{table}

Table~\ref{tab3} shows the average number of edges attacked by normal and strong jamming signals given different values of $h$ and $T$. It is interesting to note that for $h>T$, the attacker never attacks any edge with normal signals, indicating that it prefers to save its energy to use it later for more powerful attacks. Consequently, the number of edges attacked strongly with $h>T$ becomes more than those in the case of $h=T$, which results in the larger applied utilities as described above. We can also observe that in the case of $h=3$ and $T=1$, the attacker is able to strongly attack more edges than the other cases in Table~\ref{tab3} in average at $k=19$, even though at $k=9$ it attacks slightly fewer edges than the case of closer values of $h$ and $T$. This suggests that the attacker tends to save its energy more in the case of larger value of $h$ and smaller $T$.

\section{Conclusion}
We have formulated a two-player game in a cluster forming of resilient multiagent systems played over time. The players consider the impact of their actions on future communication topology and agent states, and adjust their strategies according to a rolling horizon approach. Necessary conditions and sufficient conditions for forming clusters among agents have been derived. We have discussed the effect of the weights of the utility functions and different initial states on cluster forming, and evaluated the effects of varying horizon length and game period on the players' performance.

Possible future extensions include the case where the players' utility functions are not zero-sum, the case where the players do not have perfect knowledge, and the setting where each agent is capable to decide its own strategies in a decentralized way. We have also considered in \cite{yuracc21} the case where the players' horizon lengths and game periods are not uniform. This case can be further generalized to decentralized settings where agents decide their own strategies in an asynchronous way.

{\color{black} Furthermore, it is also interesting to consider a case where the players may not have a complete knowledge of the other players. This incomplete version of the game is considered in \cite{yur23}.}

\bibliographystyle{plain}        
\bibliography{auto21}           

\appendix

\end{document}